\RequirePackage{fix-cm}

\documentclass[twocolumn,epj,nopacs]{svjour}

\usepackage[export]{adjustbox}
\usepackage{graphicx}

\usepackage{csquotes}
\usepackage[english]{babel}

\usepackage{siunitx}
\usepackage{xspace}
\usepackage[version=4]{mhchem}
\usepackage{physics}
\usepackage{xfrac}
\usepackage[switch]{lineno}
\usepackage{amsbsy}
\usepackage[artemisia]{textgreek}
\usepackage{upgreek}
\usepackage{amssymb}

\usepackage{xcolor}

\usepackage{wrapfig}

\newcommand{\hades}{HADES\xspace}
\newcommand{\fwdet}{Forward Detector\xspace}
\newcommand{\Cs}{Cross-section\xspace}
\newcommand{\Css}{Cross-sections\xspace}
\newcommand{\cs}{cross-section\xspace}
\newcommand{\css}{cross-sections\xspace}
\newcommand{\etff}{eTFF\xspace}

\newcommand{\eq}{\begin{equation}}
\newcommand{\eeq}{\end{equation}}
\newcommand{\sqs}[1][s]{\sqrt{#1}}
\newcommand{\sbs}{$S/B$\xspace}
\newcommand{\sigs}{$S$\xspace}
\newcommand{\bkgs}{$B$\xspace}
\newcommand{\signs}{$S/\sqrt{S+B}$\xspace}

\newcommand{\decays}{$\to$}
\newcommand{\Kaon}{K\xspace}
\newcommand{\aKaon}{\overline{\Kaon{}}\xspace}
\newcommand{\prot}{p\xspace}
\newcommand{\pprot}{\prot{}$_\text{p}$\xspace}
\newcommand{\deut}{d\xspace}
\newcommand{\neut}{n\xspace}
\newcommand{\nucl}{N\xspace}
\newcommand{\hype}{Y\xspace}
\newcommand{\hyped}{Y\xspace}
\newcommand{\rest}{X\xspace}
\newcommand{\nstar}{\nucl{}$^{*}$\xspace}
\newcommand{\elec}{e\xspace}
\newcommand{\vg}{\textgamma$^{*}$\xspace}
\newcommand{\photon}{\textgamma\xspace}
\newcommand{\elp}{\elec{}$^{+}$\xspace}
\newcommand{\elm}{\elec{}$^{-}$\xspace}
\newcommand{\piz}{\textpi$^{0}$\xspace}
\newcommand{\pim}{\textpi$^{-}$\xspace}
\newcommand{\pip}{\textpi$^{+}$\xspace}
\newcommand{\Kzs}{\Kaon{}$^{0}_\text{S}$\xspace}
\newcommand{\Kz}{\Kaon{}$^{0}$\xspace}
\newcommand{\Km}{\Kaon{}$^{-}$\xspace}
\newcommand{\Kp}{\Kaon{}$^{+}$\xspace}
\newcommand{\Dz}{\textDelta$^{0}$\xspace}
\newcommand{\Dp}{\textDelta$^{+}$\xspace}
\newcommand{\Dpp}{\textDelta$^{++}$\xspace}
\newcommand{\Shype}{\textSigma\xspace}
\newcommand{\Szero}{\textSigma$^0$\xspace}
\newcommand{\Splus}{\textSigma$^{+}$\xspace}
\newcommand{\Sstar}{\textSigma(1385)\xspace}

\newcommand{\Sstarz}{\textSigma(1385)$^{0}$\xspace}
\newcommand{\Sstarp}{\textSigma(1385)$^{+}$\xspace}
\newcommand{\Lzero}{\textLambda\xspace}
\newcommand{\Lstar}{\textLambda(1405)\xspace}
\newcommand{\Lstard}{\textLambda(1520)\xspace}
\newcommand{\Xim}{\textXi$^{-}$\xspace}

\newcommand{\pp}{\prot\prot}
\newcommand{\pimp}{\pim\prot}
\newcommand{\dilambda}{\Lzero\Lzero}
\newcommand{\dalitzp}{\elp\elm}

\xspaceaddexceptions{\decays \Kaon \aKaon \prot \pprot \deut \neut \nucl \hype \hyped \rest \nstar \elec \vg \photon \elp \elm \piz \pim \pip \Kzs \Kz \Km \Kp \Dz \Dpp \Shype \Szero \Splus \Sstar \starA \Sstarz \Sstarp \Lzero \Lstar \Lstard \Xim \pp \pimp \dilambda \dalitzp +}

\DeclareSIUnit\clight{\text{\ensuremath{c}}}

\DeclareSIUnit\mub{\micro\barn}

\DeclareSIUnit\gev{\giga\electronvolt}
\DeclareSIUnit\gevc{\giga\electronvolt\per\clight}
\DeclareSIUnit\gevsc{\giga\electronvolt\per\clight\squared}
\DeclareSIUnit\sgevsc{\giga\electronvolt\squared\per\clight\tothe{4}}

\DeclareSIUnit\mev{\mega\electronvolt}
\DeclareSIUnit\mevc{\mega\electronvolt\per\clight}
\DeclareSIUnit\mevsc{\mega\electronvolt\per\clight\squared}
\DeclareSIUnit\smevsc{\mega\electronvolt\squared\per\clight\tothe{4}}

\DeclareSIUnit\kev{\kilo\electronvolt}
\DeclareSIUnit\kevc{\kilo\electronvolt\per\clight}
\DeclareSIUnit\kevsc{\kilo\electronvolt\per\clight\squared}

\usepackage[inline]{enumitem}

\usepackage{tabularx}
\usepackage{booktabs}

\newcolumntype{L}[1]{>{\hsize=#1\hsize\raggedright\arraybackslash}X}%
\newcolumntype{R}[1]{>{\hsize=#1\hsize\raggedleft\arraybackslash}X}%
\newcolumntype{C}[1]{>{\hsize=#1\hsize\centering\arraybackslash}X}%

\usepackage{flushend}
\usepackage[
  backend=biber,
  style=phys,
  doi=false,
  url=false,
  isbn=false,
  eprint=true,
  giveninits=true,
  hyperref=false,
  articletitle=false,
  biblabel=brackets,%
  chaptertitle=false,
  pageranges=false,%
  maxnames=3
]{biblatex}
\DeclareFieldFormat{pages}{#1}
\DeclareFieldFormat[book]{editor}{}
\DeclareFieldFormat[phdthesis]{title}{}
\renewbibmacro{in:}{}
\bibliography{abbrv_ltwa,hades_pp45_hyperons}

\makeatletter
\def\cl@chapter{\cl@chapter \@elt {theorem}} %bug in class
\def\cl@chapter{\@elt {theorem}}
\makeatother

\usepackage[capitalize]{cleveref}

\journalname{EPJA}

\begin{document}%\linenumbers

\title{Production and electromagnetic decay of hyperons: a feasibility study with HADES as a Phase-0 experiment at FAIR}

\author{
(HADES collaboration and PANDA@HADES collaboration)\\
J.~Adamczewski-Musch\inst{4} \and
A.~Belyaev\inst{7} \and
A.~Blanco\inst{1} \and
C.~Blume\inst{8} \and
C.~Charlotte\inst{8} \and
D.S.~Borisenko\inst{12} \and
C.~Charlotte\inst{8} \and
L.~Chlad\inst{16} \and
P.~Chudoba\inst{16} \and
I.~Ciepa{\l}\inst{2}
% ...
\\
($^\dagger$ -- member of the HADES and PANDA@HADES collaborations)\\
($^\ddagger$ -- member of the PANDA and PANDA@HADES collaborations)\\
}

\institute{
LIP-Laborat\'{o}rio de Instrumenta\c{c}\~{a}o e F\'{\i}sica Experimental de Part\'{\i}culas 3004-516 Coimbra, Portugal
\and
Institute of Nuclear Physics, Polish Academy of Sciences, Krak\'{o}w, Poland
\and
Smoluchowski Institute of Physics, Jagiellonian University of Cracow, Krak\'{o}w, Poland
\and
GSI Helmholtzzentrum f\"{u}r Schwerionenforschung GmbH, Darmstadt, Germany
\and
Technische Universit\"{a}t Darmstadt, Darmstadt, Germany
\and
Institut f\"{u}r Strahlenphysik, Helmholtz-Zentrum Dresden-Rossendorf, Dresden, Germany
\and
Joint Institute of Nuclear Research, Dubna, Russia
\and
Institut f\"{u}r Kernphysik, Goethe-Universit\"{a}t, Frankfurt, Germany
\and
Physik Department E62, Technische Universit\"{a}t M\"{u}nchen, Garching, Germany
\and
II.Physikalisches Institut, Justus Liebig Universit\"{a}t Giessen, Giessen, Germany
\and
Forschungszentrum J\"ulich, J\"ulich, Germany
\and
Institute for Nuclear Research, Russian Academy of Science, Moscow, Russia
\and
Institute of Theoretical and Experimental Physics, Moscow, Russia
\and
National Research Nuclear University MEPhI (Moscow Engineering Physics Institute), Moscow, Russia
\and
Laboratoire de Physique des 2 infinis Irène Joliot-Curie, Université Paris-Saclay, CNRS-IN2P3., Orsay, France
\and
Nuclear Physics Institute, The Czech Academy of Sciences, Rez, Czech Republic
% ...
}

\authorrunning{J.~Adamczewski-Musch et al.} % if too long for running head
\titlerunning{Production and electromagnetic decay of hyperons in HADES}

\date{\today}

\abstract{
  A feasibility study has been performed in order to investigate the performance of the \hades detector to measure the electromagnetic decays of the hyperon resonances \Sstarz, \Lstar and \Lstard as well as the production of double strange baryon systems \Xim and \dilambda in p+p reactions at a beam kinetic energy of \SI{4.5}{\gev}.
  The existing \hades detector will be upgraded by a new \fwdet, which extends the detector acceptance into a range of polar angles that plays a crucial role for these investigations.
  The analysis of each channel is preceded by a consideration of the production \css. Afterwards the expected signal count rates using a target consisting of either liquid hydrogen or polyethylene are summarized.
}

\PACS{
    {PACS-key}{describing text of that key} \and
    {PACS-key}{describing text of that key}
}

\maketitle

\tableofcontents

\section{Introduction}\label{sec:intro}

Using \nucl-\nucl and \textpi-\nucl reactions,  the High Acceptance DiElectron Spectrometer (\hades \cite{Agakishiev:2009am}) explores the electromagnetic structure of baryonic resonances \cite{PhysRevC.95.065205, Ramstein:2019kaz} and investigates their production mechanisms for beam kinetic energies in the region of a few \si{\gev} \cite{PhysRevC.102.024001, Agakishiev:2017nxc, Munzer:2017hbl, Agakishiev:2015tfa}. These studies also establish important reference data for the interpretation of heavy ion induced reactions, as demonstrated in the recent measurement of dilepton emission from dense baryonic matter \cite{HADES-NATURE}. A common aspect for these reactions is the need to understand the coupling of baryons with virtual (massive) photons. These couplings can be studied in proton and pion induced reactions by measuring the Dalitz-decay of baryonic resonances (\textit{e.g.} \nstar, \textDelta) into \nucl\dalitzp. These transitions provide valuable information on electromagnetic Transition Form Factors of baryonic resonances (\etff) in the timelike kinematic region, \textit{e.g.} \textDelta(1232) \decays \nucl \dalitzp
\cite{PhysRevC.95.065205}.

In general, \etff{}s are analytical functions of the squared four-momentum transfer ($q^2$) of the virtual photon exchanged between the initial and final state baryon. They are studied in the spacelike ($q^2<0$) and the timelike ($q^2>0$) kinematic regions in electron scattering off the nucleon and in Dalitz-decays, respectively.
In the latter case, the virtual photon decays to a lepton pair (dielectron or dimuon) with an invariant mass $M$ within the region of $4(M_{l})^2<M^2<(M_\text{R}-M_\text{N})^2$, where $M_{l}$, $M_\text{R}$ and $M_\text{N}$ are the masses of the lepton, the decaying resonance and the nucleon, respectively. Note that $q^2=0$ corresponds to the radiative decay \textit{i.e.} emission of a real photon. Results published by \hades for non-strange baryons indicate a significant role in the transitions for an intermediate \textrho{} vector meson, in agreement with the Vector Meson Dominance model \cite{PhysRevC.95.065205,salabura2019}.  Microscopic calculations based on the covariant spectator quark model explain this strong coupling to the meson as predominately an effect of the pion cloud in the nucleon \cite{Ramalho:2012ng, PhysRevD.100.114014}. This conclusion complements findings from electro-scattering experiments, where meson cloud effects are important for descriptions of the eTFF in the low $q^2$ spacelike region for several baryon resonances (for an overview see \cite{Aznauryan:2011qj}).

Experimental results on hyperon radiative decays are still very sparse, despite their importance being stressed already many years ago. In the 1980's the radiative decay widths were discussed as an ideal observable to differentiate between various quark, bag and soliton models of strange baryons \cite{PhysRevD.32.695}. Since then, only a few measurements of the decays \Lstard \decays{} \Lzero \photon, \Sstarz \decays \Lzero \photon and \Sstarp \decays \Splus \photon have become available \cite{pdg, PhysRevC.71.054609, Keller:2011aw}.
The results on \Sstarz \decays \Lzero \photon decouplet-octet transitions are particularly interesting in view of the analogy within SU(3) flavor symmetry to \textDelta \decays \nucl \photon, both of which are dominated by a magnetic dipole transition \cite{Landsberg:1997mk}. The measured decay widths are found to be almost a factor 2 larger than most quark model predictions. This finding is interpreted as evidence for effects of the pion cloud.

In the timelike region, the CLEO collaboration provided the first measurement of hyperon (\Lzero, \Shype{}$^{\pm,0}$, \textXi$^{-,0}$ and \textOmega) elastic magnetic form factors in \dalitzp annihilation data at large momentum transfers of $q^2=\SI{14.2}{\gevsc\squared}$ \cite{Dobbs:2014ifa}. The form factors for the isoscalar \Lzero state are a factor \num{1.7 +- 0.2} larger than for the isovector \Szero. This has been interpreted as evidence of correlation between light quarks in the hyperon. Indeed, one expects that light quarks couple to spin $S=0$ for the \Lzero and $S=1$ for the \Szero.
Up to now, there are no measurements of \etff in Dalitz-decays of hyperons.
Since Dalitz-decays probe hyperon structure at lower $q^2$, mesonic degrees of freedom are expected to play an important role, just as for the non-strange baryons.
Moreover, this $q^2$ domain allows for a more direct connection to the spacelike region where some data are available.
Indeed, theoretical calculations based on dispersion relations \cite{Granados:2017cib} or quark models \cite{Ramalho:2011pp, PhysRevD.102.054016} provide predictions for decouplet-octet transitions in both the timelike and spacelike regions.
In particular, recent calculations on \Sstarz\decays \Lzero\dalitzp show, in analogy to \textDelta\decays \nucl\dalitzp, significant effects of the pion cloud in the distribution of dilepton invariant mass \cite{PhysRevD.102.054016}.
As mentioned above, pion cloud effects have already been suggested to explain the large decay width for the \textSigma$^{*0}$\xspace\decays\Lzero\photon transition \cite{Keller:2011aw}.
Also, calculations based on an effective Lagrangian of vector-baryon interactions predict significant effects of intermediate vector mesons (\textrho/\textomega/\textphi) for hyperon transitions \cite{PhysRevD.93.033004, PhysRevC.48.1318}.
Therefore, first measurements at \hades on both virtual  (\textit{i.e.} dielectron) and real photon decays: \Shype{}$^*$/\Lzero{}$^*$\decays \Lzero\dalitzp (\photon), respectively, will significantly impact our understanding of the electromagnetic structure of strange resonances in the $q^2$ region, where vector meson effects are expected to be large \cite{PhysRevC.48.1318}.

In addition, the collected data will enable detailed studies of hyperon production in proton-proton reactions and provide important reference measurements for future heavy-ion experiments exploring the high net-baryon density region of the QCD phase diagram (\textit{e.g.} FAIR, NICA, STAR-BES). In particular, essentially nothing is known about the production of hyperons with masses larger than the \Lzero and \Szero and of multi-strange hyperons in this beam energy range.
The impact of the higher lying hyperon resonances on QCD thermodynamics, on freeze-out in heavy ion collisions and on the evolution of the early universe has been recently reported \cite{alba2017workshop}.
Measurements of the respective excitation functions in \prot--A and A--A collisions are planned with the CBM detector \cite{Friman:2011zz}. HADES measurements will therefore provide a necessary reference to quantify the expected strangeness enhancement in heavy ion collisions and cold matter effects. One prominent example of the impact of such reference measurements is the puzzling enhancement of \textXi{} cascade production in \ce{ArKCl} reactions at \SI[parse-numbers=false]{1.75A}{\gev} observed by the \hades collaboration. It was shown that the measured cascade yield is higher by a factor \numrange{10}{100} than predicted in various theoretical models \cite{PhysRevLett.103.132301}.
A similar enhancement observed in the \prot+\ce{Nb} system points to effects appearing already in cold nuclear matter \cite{PhysRevLett.114.212301}. Production on correlated nucleon pairs or excitation of higher mass resonances with a significant decay branch to \textXi\Kaon\Kaon have been considered as  possible explanations  \cite{Steinheimer:2015sha, ZETENYI2018226}. This reaction is just one example illustrating the importance of the data on double-strange hyperon production in nucleon-nucleon interactions close to the production threshold. Similarly, little or no data exist on the production of higher mass ( \Sstar) or excited hyperon states  (for example \Lstard) \cite{PhysRevC.85.035203}.

An important aspect of the double-strangeness production program is double \Lzero production.
The \Lzero-\Lzero measurement is part of the more general baryon-baryon (\prot-\prot, \prot-\Lzero, \Lzero-\Lzero) correlation studies \cite{PhysRevC.99.024001}.
These data will constrain the rather poorly known hyperon-hyperon interaction, which has a key role in \Lzero-\Lzero double hypernuclei, neutron star core studies \cite{1985ApJ...293..470G} and the \Xim production mechanism.
The proposed measurement will complement upcoming studies by the PANDA collaboration of \Lzero-\Lzero and \Lzero-$\bar{\text\Lzero}$ in \prot-\prot and $\bar{\text{\prot}}$-\prot interactions, respectively \cite{doi:10.1080/10619127.2017.1351182}.

Our paper is organized as follows. \Cref{sec:spect} includes a brief overview of HADES and the \fwdet upgrade essential for the hyperon physics program outlined above, including detector simulations and test beam results.
In \cref{sec:benchmark}, we present selected benchmark channels in order to demonstrate the feasibility of \hades to measure radiative decays of hyperons and to investigate double-strangeness production.
This is followed in \cref{sec:sim} by a description of the simulation and analysis steps, starting with \Lzero reconstruction, which is a common step in all analyses presented here.
Thereafter in \cref{sec:results} the channel-specific procedures are presented.
Finally, estimated count rates are presented as well as a summary of this work and an outlook in \cref{sec:summary}.

\section{HADES and the \fwdet}\label{sec:spect}
\label{sec_det}

All measurements discussed here can be performed with proton beams provided by the SIS18 with energies up to $E_\text{kin} = \SI{4.5}{\gev}$ at \hades. \hades is a magnetic spectrometer that is in operation since 2002 at the SIS18 accelerator in the GSI Helmholtz Center for Heavy Ion Research in Darmstadt (Germany) \cite{Agakishiev:2009am}.
It measures charged hadrons (pions, kaons and protons), leptons (electrons and posi\-trons) and photons resulting from proton, secondary pion and heavy ion induced reactions on fixed proton or nuclear targets in the energy regime of a few $A\cdot$\si{\gev}.
It features an excellent invariant mass resolution for electron-positron pairs of ${\Updelta M/M} \approx \SI{2.5}{\percent}$ in the \textrho/\textomega/\textphi{} vector meson mass region.
With its versatility it is an excellent tool to study the properties of hadrons in vacuum, cold and also dense baryon-rich matter at moderate temperatures (for a recent review see \cite{Salabura:2020tou}).
\hades thus provides a lower energy reference for the future NICA and FAIR facilities and complements the region of high temperatures and small or even vanishing net baryon densities probed by experiments at SPS, RHIC and LHC.
Results obtained in the last decade by \hades showed that baryonic resonances have a fundamental role in the processes that define the physics at finite baryon density.
Since resonances are such important sources for meson production in heavy ion collisions at kinetic energies of a few \si{\gev}, these systems are often described as ``resonance matter'' \cite{doi:10.1080/10506899308221163}.

\begin{figure}[!t]
  \centering
  \includegraphics[width=1.0\linewidth]{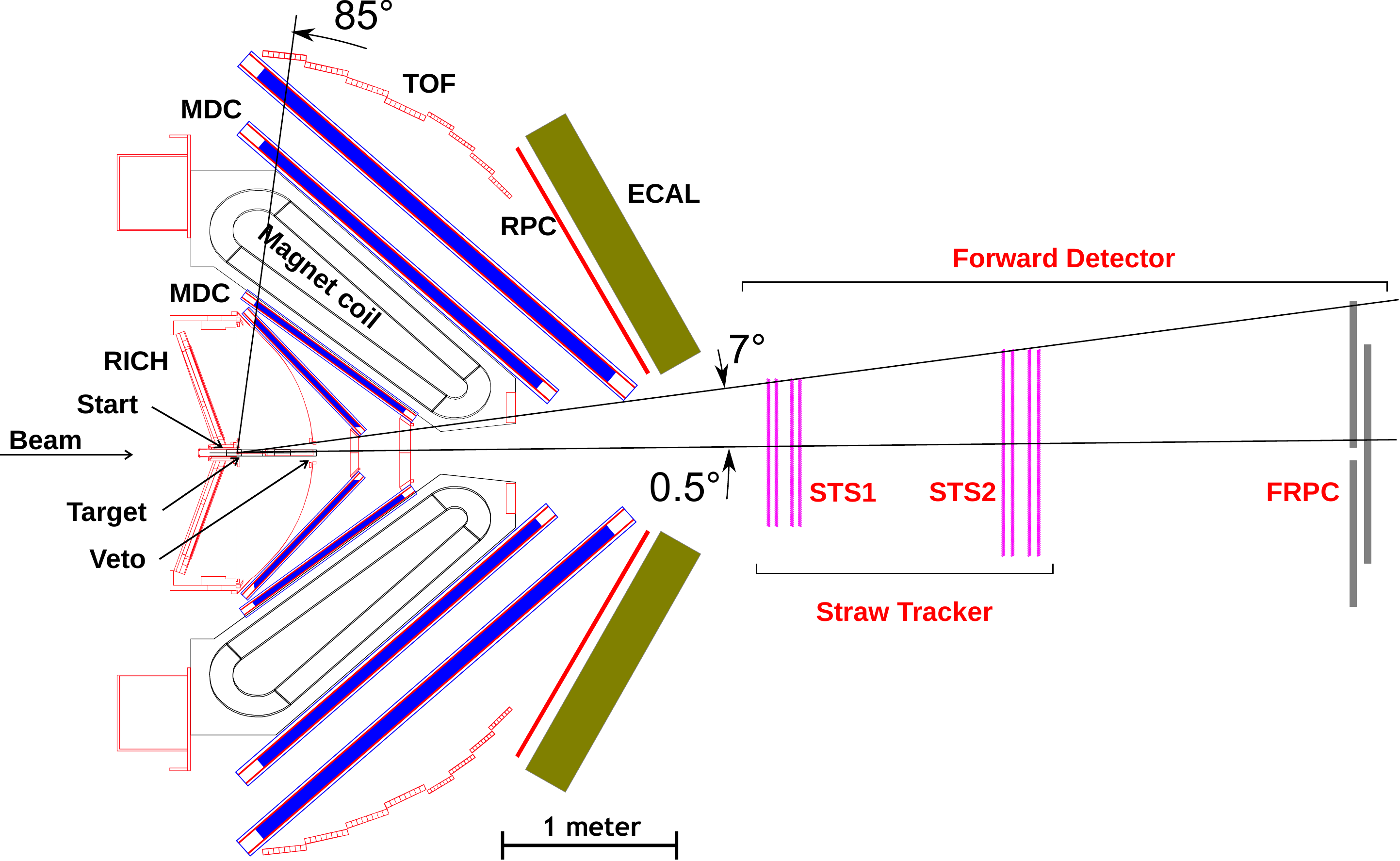}
  \caption{Schematic cross-sectional view of the \hades spectrometer. The new \fwdet extends the angular coverage into the lower polar angle range \SIrange{0.5}{7}{\degree}. The experiments described here also profit from the newly installed ECAL and upgraded RICH detectors.}
  \label{fig:fd_integration}
\end{figure}

Future \hades operation within the new experimental facility of FAIR \cite{fair}, will enable reactions to be measured with proton beam energies up to \SI{29}{\gev}, a region where string fragmentation becomes dominant.
To prepare for these new experimental challenges, the \hades spectrometer is currently undergoing several hardware upgrades: the Ring Imaging Cher\-enkov (RICH) photon detector has been upgraded to improve the dilepton identification, and an Electromagnetic Calorimeter (ECAL) has been installed for gamma reconstruction. Both components have successfully been used in 2019.
Furthermore, new and faster readout electronics (DAQ) as well as a \fwdet, to add particle tracking and velocity measurements in the forward direction, are currently being installed.
These new upgrades are expected to increase the count rate by a factor of 50--100  for many of the interesting channels.
In particular, the new setup will enable investigations of the production of hyperon states with single and double strangeness and their rare radiative decays and is a part of the FAIR Phase-0 program, in which the upgraded SIS18 accelerator is being used together with those components of the FAIR experiments that have already been completed. A schematic drawing of the upgraded setup is shown in \cref{fig:fd_integration}.

The \fwdet has full azimuthal coverage for polar angles in the range \SIrange{0.5}{7}{\degree}. It consists of two Straw Tracking Stations (STS1, STS2), which are based on instrumentation developed for PANDA \cite{Erni:2013ita, Smyrski:2018}, and a Forward Resistive Plate Chamber (FRPC) detector for Time-of-Flight (ToF) measurements. These detectors are located \SIlist{3.1;4.6;7.5}{\metre} downstream of the target, respectively. The  detectors are described in more detail below.

The maximum proton beam intensity is limited to \SI{7.5e7}{p/s} by the \hades start detector.
The existing \hades DAQ can readout a maximum trigger rate of \SI{50}{\kilo\hertz}, which is expected for this beam intensity incident on the \hades liquid hydrogen (\ce{LH_2}) target and using the envisaged trigger setting (see \cref{sec_trigger}).
These running conditions correspond to a maximum average luminosity of \SI{1.5e31}{\per \centi\meter\squared \per \second}.
The ongoing upgrade of the \hades DAQ will enable operation up to \SI{200}{\kilo\hertz}.
The increased DAQ rate combined with a more restrictive first level trigger will enable operation with a higher luminosity for the same maximum beam current by replacing the \ce{LH_2} target with a polyethylene (PE) target of the same length.
Taking the increased density of protons in the PE target ($\times$2) plus the protons in the carbon into account, the maximal luminosity for the PE target is expected to be up to a factor of $7$ higher than for the \ce{LH_2} target.
This expectation is corroborated by our results obtained from pion induced reactions, where dilepton production from the PE target is very well described by the sum of pion-proton and quasi-free pion-proton scattering \cite{ramstein2019, scozzi}.

\subsection{Straw Tracking Stations}

\begin{figure}[!t]
  \centering
  \includegraphics[width=1.0\linewidth]{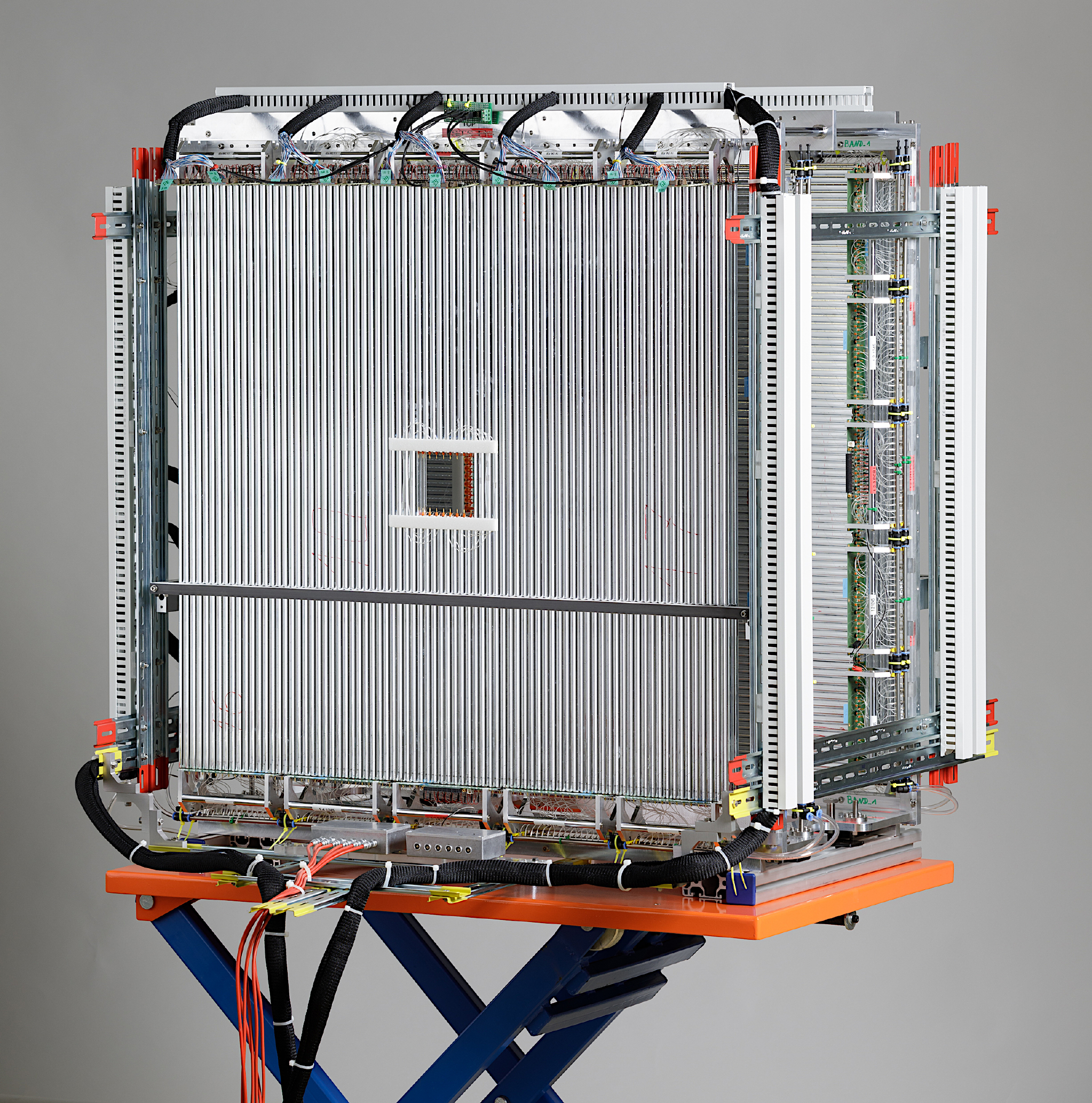}
  \caption{The STS1 prepared for installation in \hades. (Photo: courtesy of Forschungszentrum Jülich GmbH.)}
  \label{fig:sts1_drawing}
\end{figure}

The Straw Tracking Stations are based on self-supporting straw tube detectors with \SI{10}{\milli\metre} inner diameter, which were built based on the design of the Forward Tracker (FT) and the central Straw Tube Tracker (STT) of the PANDA experiment \cite{Erni:2013ita,Wintz:2014uwa,Smyrski:2017}.
The cathode of each straw consists of an aluminized Mylar foil with a thickness of \SI{27}{\micro\metre}, and the anode is a gold-plated tungsten--rhenium wire with \SI{20}{\micro\metre} diameter.
The operating pressure of the gas mixture is chosen to be \SI{1}{\bar} above the surrounding environment.
This over-pressure provides the necessary mechanical stiffness of the straws and makes them self-supporting without the need for massive support structure.
The role of the detector frames is limited to the positioning of the straws.

Groups of 32 straw tubes are glued together in two staggered layers to form a module. The modules are mounted next to each other on a common support frame, thereby forming a double-layer of straw tubes. The design of the modules is described in  Ref. \cite{Smyrski:2018} and allows for fast and simple maintenance by replacing single modules if necessary.

\begin{figure}[!t]
  \centering
  \includegraphics[width=1.0\linewidth]{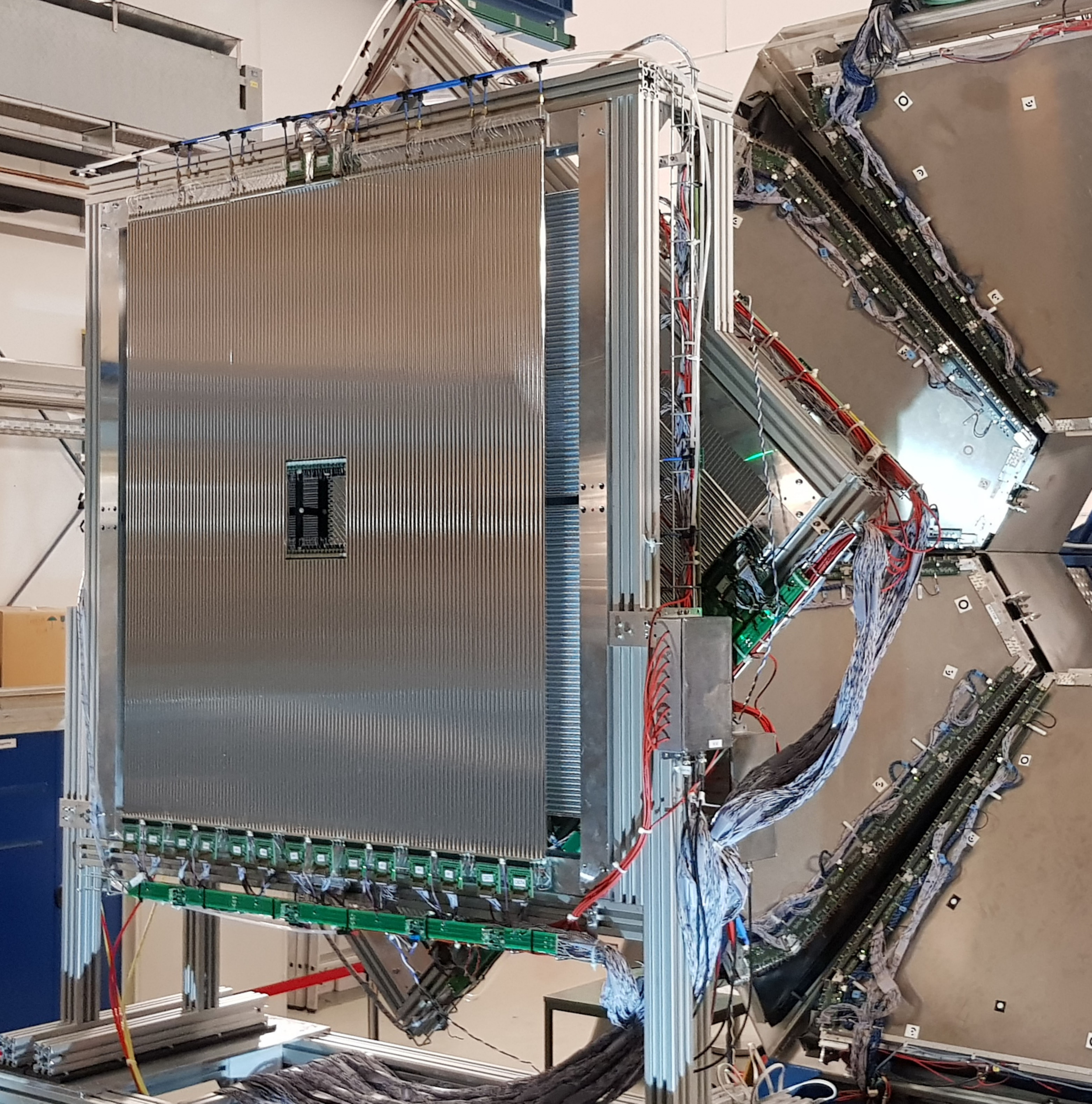}
  \caption{STS2 installed in \hades in the maintenance position. The rear planes are rotated by \ang{45} around the beam axis with respect to the front one. The equipment behind the STS2 is part of the ECAL frame with the RPC detectors in front of it.}
  \label{fig:sts2_drawing}
\end{figure}

The STS1 and STS2 each consists of four double-layers.
Groups of two double-layers are mounted on an individual support structure with one double-layer on each side of the support.
The double-layers of straws are aligned by an inclination of \SIlist{0;90;90;0}{\degree} with respect to the vertical direction in consecutive double-layers of the STS1 (see \cref{fig:sts1_drawing}). In the STS2 the straws in the first two double-layers are inclined by \ang{90} and \ang{0} and the next double-layers are inclined by \ang[retain-explicit-plus]{+45} and \ang{-45} (see \cref{fig:sts2_drawing}). This configuration facilitates an unambiguous reconstruction of multi-track events.
The straws in the STS1 and STS2 have a length of \SI{76}{cm} and \SI{125}{cm}, respectively. The double-layers are \SI{80}{cm} and \SI{112}{cm} wide, respectively.
The double-layers in STS1 and STS2 contain a central opening for the beam of \SI{8 x 8}{\cm} and \SI{16 x 16}{\cm}, respectively.
The opening is created by replacing the most central straws in the double-layers with pairs of shorter straws, leaving a gap between them.
In total, the STS1 and STS2 stations contain 704 and 1024 individual straw tubes, respectively.

The pulses from the anode wires are amplified and discriminated in front-end electronic cards based on the PASTTREC ASIC \cite{Przyborowski:2016}. Both the leading-edge time and the time-over-threshold of the detector signals are measured in multi-channel TDCs implemented in the Trigger Readout Board version 3 (TRB3) \cite{Traxler:2011}.

The straw tubes are operated with a gas mixture of \ce{Ar}\ce{CO_2} (90:10) at \SI{2}{\bar} absolute pressure.
The operating voltage of the anode wires is \SI{1800}{\volt}, resulting in a gas gain of about \num{5e4} and a maximum drift time of about \SI{130}{ns} for a maximum drift length equal to the straw tube radius (\SI{5}{mm}).
The spatial resolution for each straw tube layer has been measured to be about \SI{0.13}{mm} ($\sigma$) for minimum ionizing protons \cite{Smyrski:2018}.
The average hit detection efficiency for protons with momentum of \SI{2}{\gevc} incident on a single straw is measured to be over \SI{95}{\percent}.
The measured time-over-threshold of the detector signals is used to reject noise pulses and can also be used to determine the specific energy loss in order to distinguish protons, charged pions and kaons with momentum up to about \SI{0.8}{\gevc} \cite{Jowzaee:2013, Strzempek:2017}.

\subsection{The FRPC Time-of-Flight detector}

The FRPC is a ToF system in the \fwdet based on individually shielded resistive plate chambers  \cite{FINCK200363} and follows the existing concept in the \hades ToF wall \cite{HADES}, see \cref{fig:fd_integration}.

\begin{figure}[!t]
  \centering
  \includegraphics[width=1.0\linewidth]{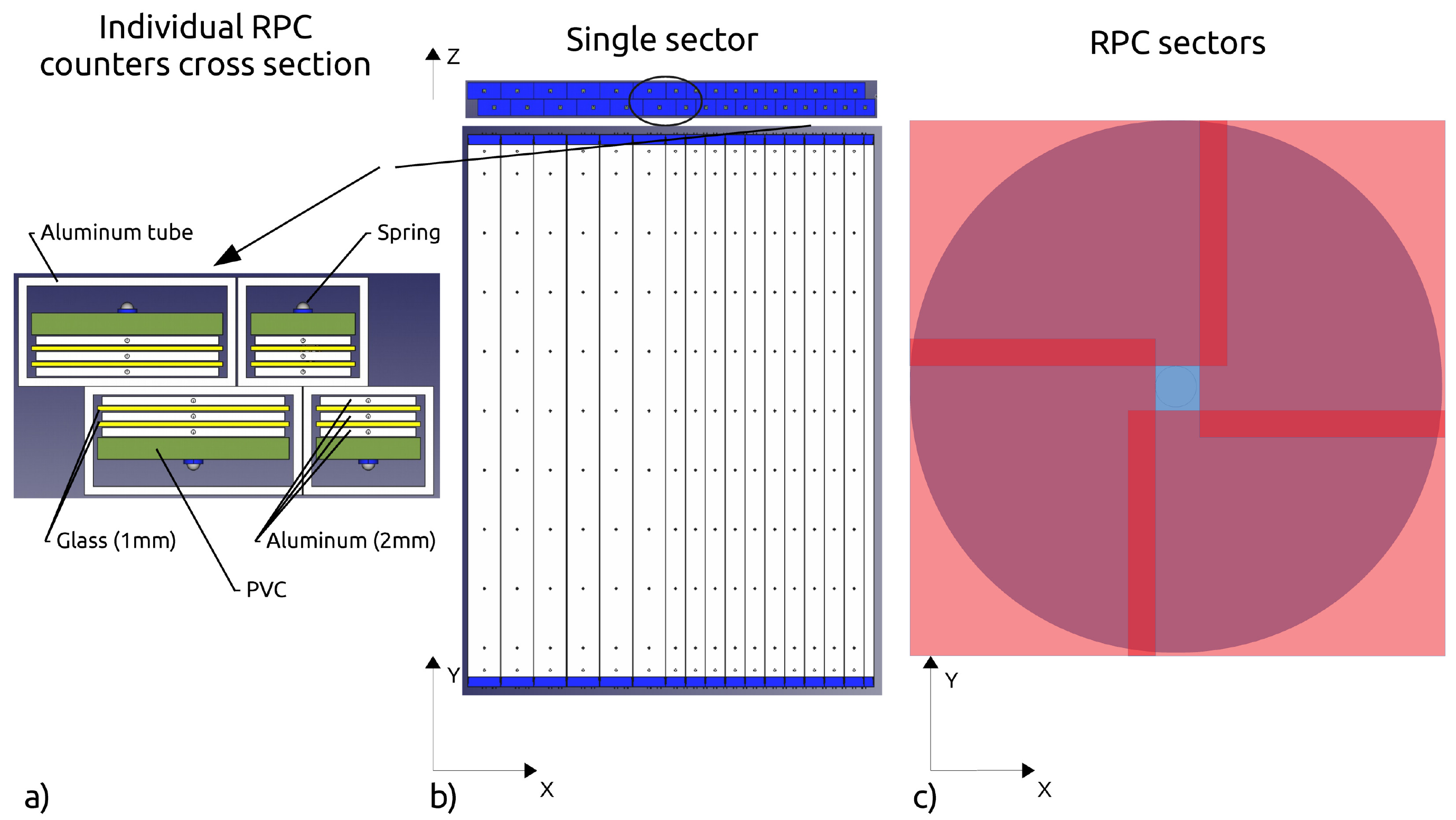}
  \caption{Drawing of the FRPC. a) Cross sectional view of two \SI{22}{\milli\metre} and two \SI{42}{\milli\metre} width individual RPC counters, b) single sector with two partially overlapping layers consisting of sixteen individual RPC counters each and c) arrangement of the four RPC sectors.}
  \label{fig:rpc_drawing}
\end{figure}

The FRPC detector consists of individually shielded hybrid (metal glass) strip-like RPC counters, which consist of three aluminum electrodes (\SI{2}{\milli\metre} thick) and two glass electrodes (\SI{1}{\milli\metre} thick), see \cref{fig:rpc_drawing}~a).
The four gaps are defined by PEEK (Polyetheretherketone) mono-filaments that have \SI{270}{\micro\metre} diameter and are spac\-ed every \SI{90}{\milli\metre} along the counter.
The assembly is housed inside individual shielding tubes and compressed with controlled force by springs. The springs act on the Polyvinyl chloride (PVC) plate, which distributes the force.
High-voltage (HV) is applied to the central aluminum electrode, while the outer electrodes are grounded and the glass electrodes are kept electrically floating.
Insulation to the shielding tube walls is assured by a double-layer KAPTON\texttrademark{} adhesive laminate. An end-shield made of aluminum foil is spot-welded on the shielding tubes.
The signals are collected in the central electrode, at both ends, through a \SI{2}{\nano\farad} HV coupling capacitor and transported outside the gas box through \SI{50}{\ohm} cables and MMCX feedthroughs, that cross the lateral wall of the gas box.
Signals are amplified by low-jitter high-gain/bandwidth Front-End Electronics (FEE) \cite{Belver2010} and the ToF as well as the time-over-threshold of the detector signals is measured using the TRB3.

The counters are arranged in sectors, each with two partially overlapping layers, (see \cref{fig:rpc_drawing}~b).
Each layer is composed of sixteen counters \SI{750}{\milli\metre} long and two different widths of \SIlist{22;42}{\milli\metre} in order to accommodate the spatial variation of the particle flux.
The complete ToF detector is composed of four sectors arranged as shown in \cref{fig:rpc_drawing}~c), providing a hole of \SI{130}{\milli\metre} diameter for the beam to pass through.
The whole device is located at \SI{7.5}{\metre} downstream of the target.

The ToF detector for the \fwdet must be capable of measuring particle rates of \SI{320}{\hertz\per\centi\metre\squared} with an efficiency better than \SI{90}{\percent} and a time resolution better than \SI{100}{\pico\second}.
A test of a first prototype has been performed at the COSY accelerator with a beam of \SI{2.2}{\gevc} protons.
The obtained results show that the detector is able to fulfill these specifications for particle fluxes up to \SI{250}{\hertz\per\centi\metre\squared} at room temperature.
The glass resistivity drops by a factor 10 for an increase of the temperature by \SI{25}{\celsius}, enabling a corresponding increase of the FRPC rate capability \cite{GONZALEZDIAZ200572} by increasing the operational temperature of the detector.
A rate of \SI{500}{\hertz\per\centi\metre\squared} was obtained at \SI{40}{\celsius}, with similar efficiency and resolution.
A sustained rate of \SI{1}{\kilo\hertz\per\centi\metre\squared} was obtained with a modest drop of the detection efficiency down to \SI{89}{\percent}.

In view of the need to increase the operating temperature of the detector to achieve (or exceed) the count rate requirements, the whole sector is made out of heat-resistant materials, chosen to withstand temperatures up to \SI{65}{\celsius}, without loss of the relevant properties, and is equipped with a heating system.

\subsection{Digitization and track reconstruction}

\begin{figure}[!b]
  \centering
  \includegraphics[width=1.0\linewidth,valign=m]{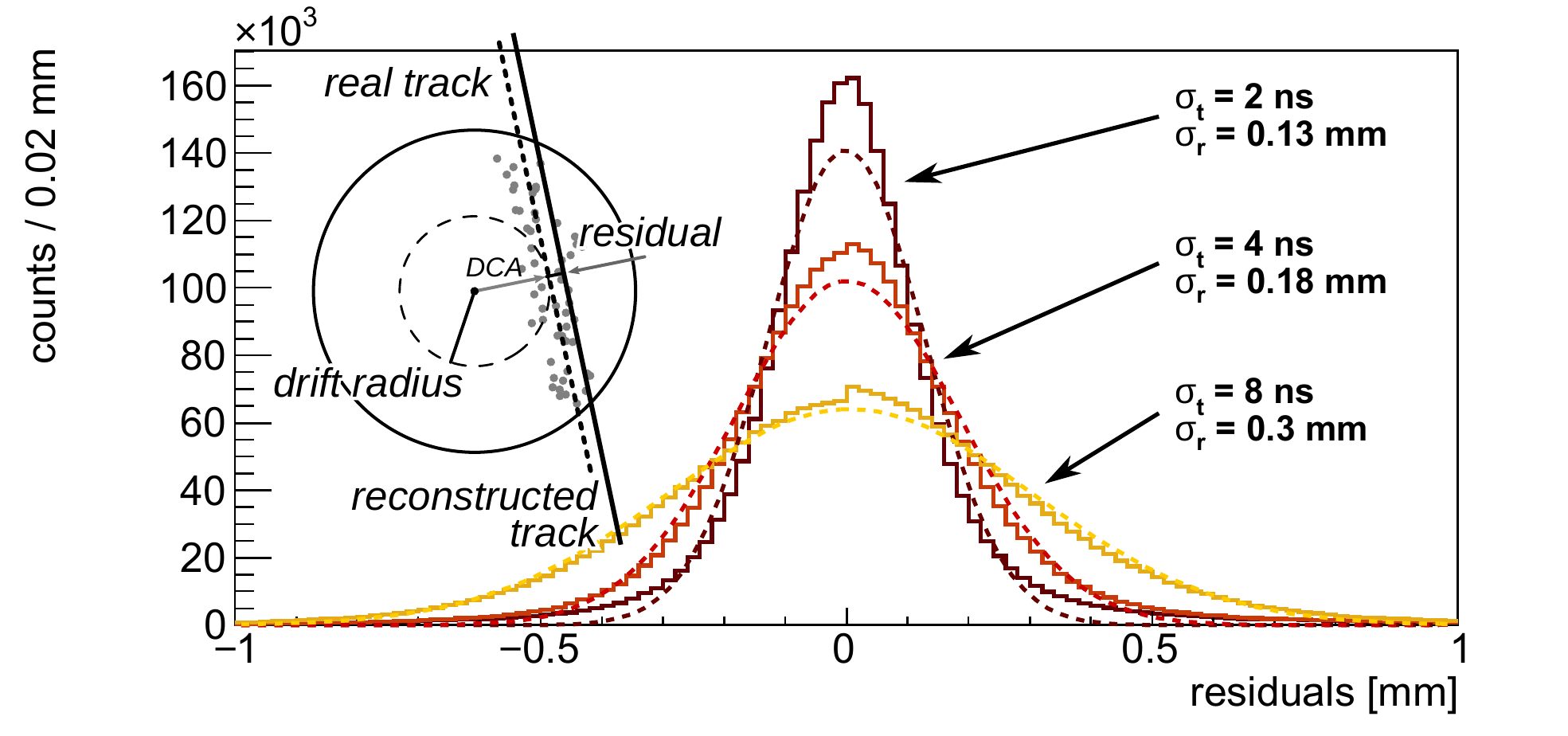}

  \caption{Influence of the drift time smearing chosen in the simulations on the calculated track residual. The solid color distributions corresponds to residuals for different time resolutions, respective dotted lines are gaussian fits.
    (inlet) Charged particle passing the straws create electron--positive ion pairs. Electrons drift towards the anode and the drift radius is calculated from the drift time. The radius represents the distance of closest approach (DCA) between the particle track and the anode. The difference between the calculated drift radius and the reconstructed track is called the residual.}
  \label{fig:sts_residual}
\end{figure}

Each individual straw is implemented in the \hades simulation framework taking into account the size and material of the straw film, gas volume and the anode wire.
In the GEANT3 \cite{Brun:118715} based simulations the particle entrance and exit points of the tube are registered in order to approximate the track passing through the tube.
For each tube with a registered hit, the digitization consists of first determining the track path by interpolating between the entrance and exit points with a line segment and then calculating the minimum distance to the center of the tube (\textit{i.e.} neglecting the wire thickness) (see \cref{fig:sts_residual}-inlet in the left).
Then the track-to-wire distance is converted into the corresponding drift time by inverting the measured drift time spectrum \cite{Apostolou:2018gac}.
In order to account for effects such as the time resolution of the front-end electronics, possible gain variations, temperature and systematic errors in the detector geometry, the drift time is convoluted with a Gaussian distribution with $\sigma=\SI{4}{\nano\second}$ \cite{Strzempek:2017}. The finite detector inefficiency is accounted for by only storing the hit with \SI{95}{\percent} probability.

The track reconstruction proceeds in two steps: First a low-resolution (LR) tracking procedure is applied to the data. In this process only the straw anode coordinates are used, thereby limiting the tracking precision to the pitch of the straw detector anodes.
Then the track is matched to a hit in the FRPC.
The timing signal from the FRPC is used to calculate the track-to-wire distances from the drift time in the individual straw detectors, thereby enabling a high-resolution (HR) tracking mode.

\begin{figure}[!t]
  \centering
  \includegraphics[width=0.8\linewidth]{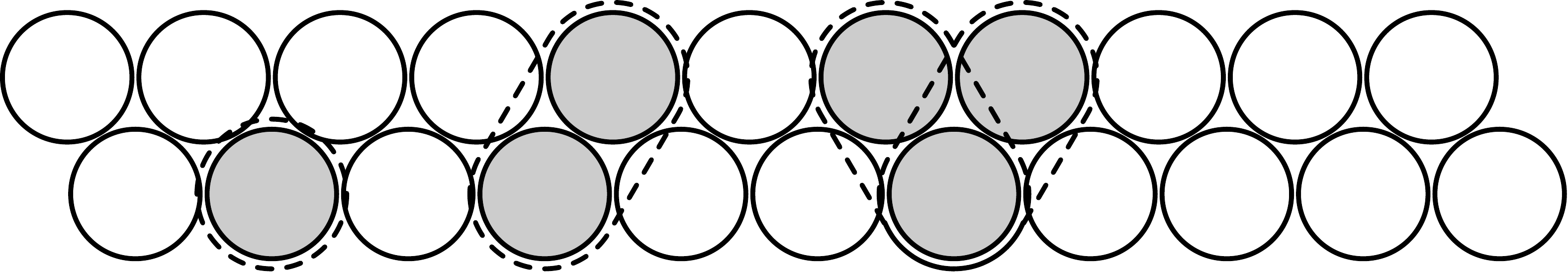}
  \caption{Each circle represents an individual straw, and the fired straws are marked by gray filled circles. The dashed outlines show four possible types of clusters.}
  \label{fig:sts_clusters}
\end{figure}

In the tracking algorithm, all registered hits from the same straw double-layer are grouped into clusters of one or two hits (neighboring channels, see \cref{fig:sts_clusters}). Successful tracking requires that there is a cluster in each double-layer of straws, thus the detection probability in a double-layer is $P = 1 - 0.05^{2} = 0.9975$. All combinations of clusters from all layers are fit with straight lines. Wrong combinations are rejected based on the $\chi^2$ value determined assuming a spatial resolution of $\sigma = \SI{3}{\milli\metre} \approx \SI{10}{\milli\metre}/\sqrt{12}$. Only the best track candidates proceed through to the further tracking steps.

Two track fitting methods have been tested for the \fwdet:
\begin{enumerate*}[label=\emph{\alph*)}]
  \item a Least Squares Method (LSM) adopted from the CBM-MUCH detector \cite{ZinchenkoCBM2014}, and
  \item a $\chi^2$ minimization adopted from the COSY-TOF detector \cite{Jowzaee:2015qda}.
\end{enumerate*}
The LSM uses an equation solving computation, whereas the $\chi^2$ minimization uses Minuit. The $\chi^2$ minimization method offers a more general solution, because LSM does not allow a track to be perpendicular to the beam axis (this case however does not allow hits in multiple double-layers due to the STS detector geometry). Track candidates reconstructed by the HR procedure are selected based on the $\chi^2$ value using a tracking resolution of $\sigma = \SI{200}{\micro\metre}$ for each layer. In the final stage, tracks are sorted by their $\chi^2$ value from lowest to highest. Each pair of tracks which share at least half of their straws is considered and the track with the larger $\chi^2$ value is rejected. The track momentum is calculated using the track length and the FRPC ToF and assuming that the particle is a proton.
In this case, the relative proton momentum resolution varies nearly linearly from $\delta p/p$ (FWHM) = \SI{0.5}{\percent} at \SI{0.5}{\gevc} to \SI{3}{\percent} at \SI{2}{\gevc}.
Extrapolation of single tracks from the \fwdet back to the target passes within \SI{2.3}{\mm} ($\sigma$) of the true vertex. The combination of two tracks to a hyperon decay vertex is measured with a precision of $\sigma_{x,y} = $\SI{3}{\mm}, $\sigma_{z} = $\SI{11.2}{\mm} if one track is in the \fwdet. If both tracks are in \hades the vertex resolution is $\sigma_{x,y} = $\SI{4.2}{\mm}, $\sigma_{z} = $\SI{9.4}{\mm}. The resolution of both the single track and the hyperon vertex is primarily limited by straggling in the air between the target and the \fwdet, and does not significantly depend upon the drift time resolution of an individual straw, within the range \SIlist{2;4;8}{\ns} (\cref{fig:sts_residual}).

Results from tests of the straw detector exposed to a beam of \SI{2}{\gevc} protons from the COSY accelerator at FZ-J\"{u}lich were compared to simulations.
A test setup consisting of 16 double-layers arranged according to the PANDA Forward Tracker geometry: eight straw double-layers were inclined at \SIlist[retain-explicit-plus]{0;+5;-5;0;0;+5;-5;0}{\degree}, respectively and the distance between the layers also matched the Forward Tracker.
The depth of the tracking stack was \SI{70}{\centi\metre} compared to \SI{183}{\centi\metre} for the \fwdet.
The custom tracking method used to analyze the data requires the first and the final tracking plane to always have double hit clusters.
For a given pair of clusters, a corridor with a width of five straws is defined, and the cluster in each inner tracking layer is required to be within the corridor.
The statistics of the number of fired straws in the inner layers (maximum 12 straws out of 16) follows the binomial distribution and yields a single straw efficiency of about \SIrange{95}{97}{\percent} at \SI{1750}{\volt}.
The applied tracking procedure considered only eight double-layers with vertical straws \ang{0} layers, and the remaining eight inclined layers were not used
in order to more closely resemble the conditions of the Forward Detector.
For the distribution of residuals, a standard deviation of \SI{156}{\micro\metre} was obtained, which is consistent with the assumed drift time smearing of \SI{4}{\ns}.

\section{Benchmark channels}\label{sec:benchmark}

\begin{table}[!b]
  \caption{Production \cs for the signal channels used in the benchmark simulations. See \cref{sec:xs_est} for details.}
  \label{tab:sig_cs}
  \begin{tabularx}{1.0\linewidth}{L{0.4}C{0.3}C{0.3}}
    \toprule
    Channel         & $\sigma$ ($\si{\micro\barn}$) & Reaction Group           \\
    \midrule
    \Sstarz\rest    & 56.2                          & AB\phantom{CD}           \\
    \Lstar \rest    & 32.2                          & AB\phantom{CD}           \\
    \Lstard\rest    & 69.6                          & AB\phantom{CD}           \\
    \prot\Kp\Kp\Xim & 0.35-3.6                      & \leavevmode\hphantom{AB}C\hphantom{D} \\
    \dilambda\Kp\Kp & 0.35-3.6                      & \leavevmode\phantom{ABC}D           \\
    \bottomrule
  \end{tabularx}
\end{table}

Several groups of benchmark channels have been chosen in order to investigate the performance of the upgraded \hades detector to measure the following processes in proton-proton collisions at a beam kinetic energy of $E_\text{kin}=\SI{4.5}{\gev}$:
\begin{enumerate}[label={\Alph*})]
  \item real photon electromagnetic decay of hyperons,
  \item virtual photon electromagnetic decay of hyperons,
  \item double strangeness (\Xim) production,
  \item double \Lzero production and their correlations.
\end{enumerate}

A full list of the production \css for the signal channels studied in this work is provided in \cref{tab:sig_cs}. The corresponding decay branching ratios are listed in \cref{tab:sig_br}.
Except for the channels with real photon decay, the signal reconstruction strategy assumes a semi-inclusive reconstruction tagged by the \Lzero \decays \prot\pim weak decay with hyperons reconstructed using the invariant mass of their decay products.
This strategy takes advantage of both the larger acceptance for inclusive reconstruction in \hades and the larger inclusive \css.
However, the real photon decay of hyperons requires exclusive reconstruction in order to reject background stemming from neutral pion decays.
The dominant background channels included in the simulations are listed in \cref{tab:bkg_cs}.
They can be grouped into reactions with final states including:
\begin{enumerate*}[label=\emph{(\alph*})]
  \item at least one proton and multi-pions (with at least one negative charge)
  \item associated strangeness production with \Lzero/\Szero hyperons, and
  \item non-strange resonances \textit{e.g.} \textDelta(1232), relevant for the studies of Dalitz-decays.
\end{enumerate*}

\subsection{\Cs estimates}\label{sec:xs_est}

No data on the inclusive production \css of the hyperon resonances \Sstarz, \Lstar, \Lstard (denoted in the article as \hyped) and \Xim hyperons exist for proton-proton reactions in the energy region of interest. Thus, the following \cs estimates are based on the available data on \Lzero/\Szero production.

\begin{figure}[!t]
  \centering
  \includegraphics[width=1.0\linewidth]{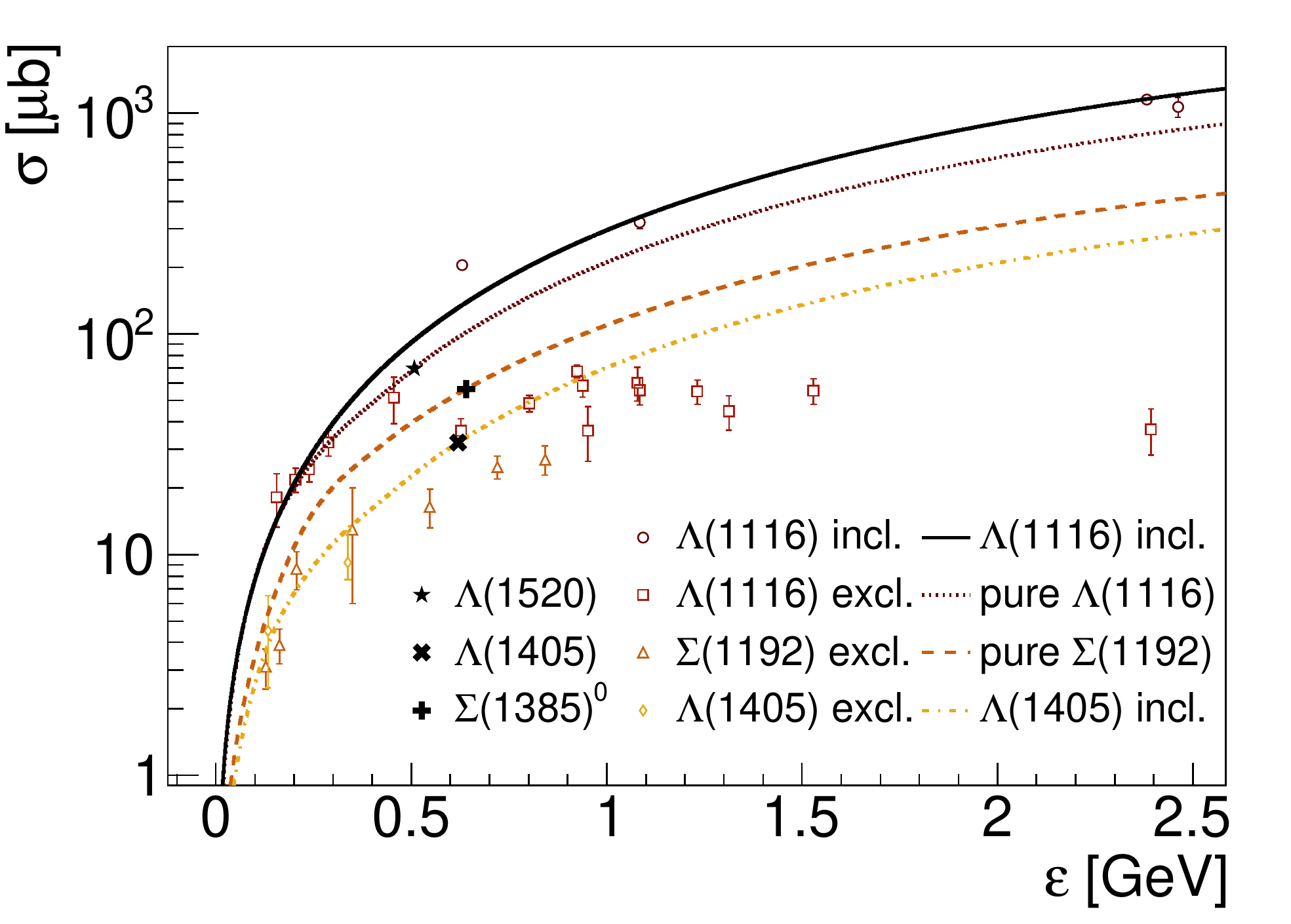}
  \caption{Hyperon production \cs as a function of excess energy $\epsilon$. The solid black line represents a parameterization of $\sigma_\text{\pp\decays\Lzero\rest}$ (see text for details). The dotted and dashed lines represent a decomposition of the inclusive \Lzero \cs into \Lzero corrected for the \Szero feed-down (dotted) and \Szero (dashed) contributions. The short-dashed line is a parameterization of $\sigma_\text{\pp\decays\Lstar\rest}$, obtained as $1/3$ of $\sigma_\text{\pp\decays\Lzero\rest}$ (see text for details). The black full symbols represent the estimated inclusive \css for \Lstard, \Lstar and \Sstarz for \pp reactions with $E_\text{kin}=\SI{4.5}{\gev}$. The open symbols represent existing experimental data.}
  \label{fig:CS_all_chanells}
\end{figure}

\subsubsection{Production \css of excited hyperons}

Inclusive production of hyperon resonances (see \cref{tab:sig_cs}) is approximated in simulation by the most dominant reaction channel of the type \prot\Kp\hyped with \css calculated as explained below.

\Css for the exclusive reaction \pp \decays \prot\Kp\Lzero have been measured in the range $\SI{2.5}{\gev} < \sqs < \SI{6}{\gev}$ by several experiments: COSY-TOF \cite{COSY-TOF_EPJA}, \hades \cite{PhysRevC.95.015207} and older ones summarized in \cite{L-B} (see \cref{fig:CS_all_chanells}).
Data on the inclusive \cs for \pp \decays \Lzero\rest are sparse; there are only four data points in the energy range of interest. The inclusive \Lzero excitation function is described by using a third order polynomial in order to interpolate these data points and the exclusive data for energies close to the production threshold. The interpolation function, denoted by the solid line is:
\begin{equation}
  \label{eq:lambda0cs}
  \sigma_{\text{\pp\decays\Lzero} \rest} = 47.97 \epsilon+292.6 \epsilon^2 -45.36 \epsilon^3\ \si{\mub}
\end{equation}
with the excess energy $\epsilon = \sqrt{s}-\sqrt{s_{th}}$, where $\sqrt{s_{th}} = \SI{2.55}{\gev}$ is the threshold for \Lzero production.

Since the branching ratio for \Szero into \Lzero\photon
is almost \SI{100}{\percent}, the inclusive \Lzero \cs includes the feed-down from \Szero production. Nevertheless, it is possible to disentangle these two channels using the measured \Lzero/\Szero ratio in exclusive channels, as given in \cite{ISI:000435653100097}. Thus, the measured \Lzero/\Szero \cs ratio was parameterized by the COSY parameterization \cite{COSY-TOF_EPJA} for $\epsilon < \SI{0.275}{\gev}$ and a linear function
\begin{equation}
  \frac{\sigma_\text{\Lzero\rest}}{\sigma_\text{\Szero\rest}}(\epsilon)
    = \num{2.215} - \num{0.027} \epsilon
\end{equation}
to describe the ratio for $\epsilon>\SI{0.275}{\gev}$. The parameterization of the inclusive \Lzero production \cs, corrected for \Szero feed-down, is shown in \cref{fig:CS_all_chanells} by the dotted line. The corresponding parameterization for the \Szero is plotted with a dashed line.

The \cs for \pp \decays \Lstard \rest is estimated assuming the same matrix element as for the \pp \decays \Lzero \rest reaction, after correction for the \Szero feed-down. Consequently, it was assumed that the only difference between the \css for these two channels at a given $\sqs$ is the volume of available phase-space. The respective threshold energy to produce the \Lstard is
$\sqs[s_\text{thr}] = \SI{2.95}{\gev}$, thus
at $E_\text{kin}=$ \SI{4.5}{\gev} ($\sqs=\SI{3.46}{\gev}$) the excess energy corresponds to $\epsilon = \SI{0.51}{\gev}$. At this $\epsilon$ and assuming
\begin{equation}
  \sigma_\text{\Lzero\rest}(\sqs=2.55+\epsilon) \approx
  \sigma_\text{\Lstard\rest}(\sqs=2.95+\epsilon)
\end{equation}
a value of $\sigma_\text{\Lstard\rest} = \SI{69.6}{\micro\barn}$ is obtained.

The \cs for inclusive \Sstarz production was calculated in a similar way as for the \Lstard. The input parameterization obtained for the \Szero was used under the assumption that the difference between the \Sstar and the \Szero production at the given $\sqs$ is governed only by the volume of available phase-space. The respective \cs for the inclusive production at $E_\text{kin}=\SI{4.5}{\gev}$ has been estimated in this way to be $\sigma_\text{\Sstar\rest}=\SI{56.5}{\micro\barn}$.

\begin{table}[!b]
  \caption{List of branching ratios for the signal and background channels, see \cref{sec:xs_est} for details. The "Reaction Group" column denotes the reaction from \cref{tab:sig_cs} for which this BR is used.}
  \label{tab:sig_br}
  \begin{tabularx}{1.0\linewidth}{L{0.4}L{0.3}C{0.3}}
    \toprule
    Channel                        & Branching Ratio & Reaction Group           \\
    \midrule
    \Sstarz \decays \Lzero\photon  & \num{1.4e-2}    & A\phantom{BCD}           \\
    \Sstarz \decays \Lzero\dalitzp & \num{1.4e-4}    & \leavevmode\phantom{A}B\phantom{CD} \\
    \Lstar  \decays \Lzero\photon  & \num{5.0e-4}    & A\phantom{BCD}           \\
    \Lstar  \decays \Lzero\dalitzp & \num{4.9e-6}    & \leavevmode\phantom{A}B\phantom{CD} \\
    \Lstard \decays \Lzero\photon  & \num{1.1e-2}    & A\phantom{BCD}           \\
    \Lstard \decays \Lzero\dalitzp & \num{1.1e-4}    & \leavevmode\phantom{A}B\phantom{CD} \\
    \Dp     \decays \prot\elp\elm  & \num{4.5e-5}    & \leavevmode\phantom{A}B\phantom{CD} \\
    \Dz     \decays \neut\elp\elm  & \num{4.5e-5}    & \leavevmode\phantom{A}B\phantom{CD} \\
    \bottomrule
  \end{tabularx}
\end{table}

The \cs for the exclusive \Lstar production was measured by ANKE \cite{ANKE_L1405} and \hades \cite{PhysRevC.87.025201}.
An empirical relation between the \cs as a function of the excess energy for \Lstar and \Lzero was deduced from these data to be: $\frac{1}{3}\sigma_\text{\Lzero\Kp\prot}(\epsilon) = \sigma_\text{\Lstar\Kp\prot}(\epsilon)$ in \cite{PhysRevC.87.025201}. Using this relation and the parameterization for inclusive \Lzero production, the inclusive \cs for \Lstar production is estimated to be $\SI{32}{\micro\barn}$ at $\sqs=\SI{3.46}{\gev}$. The results are summarized in \cref{tab:sig_cs}.

\subsubsection{Decay branching ratios of excited hyperons} \label{sec:sig_br}

The branching ratios (BR) for the real photon radiative transition of \Sstarz, \Lstard were obtained from CLAS \cite{PhysRevC.71.054609, Keller:2011aw} and are given in \cref{tab:sig_br}.
The branching ratios for the Dalitz-decay of these states have been estimated from the measured branching ratio for real photon decay using the following formula \cite{PhysRevD.93.033004}:
\begin{equation}
  \dv{\Gamma^\text{\hyped\decays\Lzero\dalitzp}}{M_\text{ee}} \simeq
  \frac{2\alpha}{3\pi{}M_\text{ee}}\Gamma^\text{\hyped\decays\Lzero\photon} \text{,}
\end{equation}
where $\alpha$ is the fine structure constant and $M_\text{ee}$ is the dilepton invariant mass. After integration over $M_\text{ee}$ this simplifies to
\begin{equation}
  \Gamma^\text{\hyped\decays\Lzero\dalitzp}=1.35 \cdot \alpha \Gamma^\text{\hyped\decays\Lzero\photon}\text{.}
\end{equation}

\subsubsection{Production of double strangeness}

The \cs for \pp\decays\prot\Kp\Kp\Xim at $\sqs=\SI{3.46}{\gev}$ ($\epsilon=\SI{0.2}{\gev}$) has been estimated from the \cs ratio $\sigma_{\text\Xim\rest}/$ $\sigma_{(\text{\Lzero+\Szero})\rest} = [\num{1.2}\num{+-0.3}(\text{stat})\num{+-0.4}(\text{syst})]\times\num{e-2}$ measured in proton-nucleus interactions at $E_\text{kin}=\SI{3.5}{\gev}$ by \hades \cite{PhysRevLett.114.212301}.
The \cs for inclusive \Lzero production in proton-proton collisions (uncorrected for the \Szero feed-down) has been parameterized as described above.
At $E_\text{kin}=\SI{4.5}{\gev}$ the \Xim production  \cs is thus estimated to be \SI{3.56}{\micro\barn}.

This estimate is based on the measured ratio $\sigma_{\text\Xim\rest}/\sigma_{(\text{\Lzero+\Szero})\rest}$ in proton-nucleus interactions, which might be strongly affected by medium effects.
In fact, the measured ratio is surprisingly large and exceeds predictions from the UrQMD transport model \cite{PhysRevLett.114.212301}.
Possible explanations of this puzzling result are given in \cite{Steinheimer:2015sha}, including decays of higher mass non-strange resonances with sizeable branching ratios into \textXi\Kaon\Kaon.
If such a mechanism is significant, it should also be active in proton-proton interactions.
Hence, this estimate is considered as an upper limit for the production \cs.
In order to make an alternative estimate, the \cs ratio $\sigma_\text{\Xim\rest}/$ $\sigma_{(\text{\Lzero+\Szero})\rest}$ at the same excess energy has been calculated from existing \pp data.
The lowest energy for which the \cs of \pp \decays \Xim\Kaon\Kaon\rest was measured ($\sigma=\SI{7}{\mub}$) corresponds to $\sqs = \SI{4.54}{\gev}$ or $\epsilon=\SI{1.29}{\gev}$ \cite{L-B}. At this excess energy the inclusive \cs for \Lzero production is $\sigma_{\text{\Lzero\rest}}=\SI{451}{\micro\barn}$ as estimated from \cref{eq:lambda0cs}. Using these data, the ratio $\sigma_{\text{\text\Xim\rest}}/\sigma_{(\text{\Lzero+\Szero})\rest}$ is 0.015.
Extrapolating this ratio to $E_\text{kin}=\SI{4.5}{\gev}$ results in a value of $\sigma_\text{\Xim\rest}=\SI{0.35}{\micro\barn}$, which is taken as an alternative estimate of the production \cs in these simulations.

The exclusive production \cs for \dilambda ($\sqs[s_{th}] = \SI{3.21}{\gev}$) was taken to be equal to the \cs for \pp \decays{} \prot\Kp\Kp\Xim ($\sqs[s_{th}] = \SI{3.236}{\gev}$). This is motivated by the same strange quark content, meson number, and the minimal difference in the volume of available phase-space due to the different threshold energies of about \SI{26}{\mev}.

\subsubsection{Background channels}\label{sec:bg_channels}

A list of the dominant background channels included in these simulations is presented in \Cref{tab:bkg_cs}. These background channels contain the same final state particles as for the signal channels, \textit{e.g.} \prot\pim\dalitzp for hyperon Dalitz-decays, \prot\pim\pim for \Xim, and \prot\prot\pim\pim for \dilambda.

\begin{table}[!t]
  \caption{List of background channels used in the benchmark simulations. See \cref{sec:xs_est} for details. The "Reaction Group" column denotes the reaction in \cref{tab:sig_cs} for which the given channel is used.}
  \label{tab:bkg_cs}
  \begin{tabularx}{1.0\linewidth}{R{0.05}L{0.4}L{0.3}C{0.3}}
    \toprule
       & Channel                    & $\sigma$ ($\si{\micro\barn}$) & Reaction Group           \\
    \midrule

    1  & \prot\prot\pip\pim\piz     & 1840                          & AB\phantom{AB}           \\
    2  & \prot\pip\pim\Dp           & 2760                          & \leavevmode\phantom{A}B\phantom{CD} \\
    3  & \prot\neut\pip\pip\pim\piz & \leavevmode\phantom{0}300     & \leavevmode\phantom{A}B\phantom{CD} \\
    4  & \prot\pip\pip\pim\Dz       & \leavevmode\phantom{0}450     & \leavevmode\phantom{A}B\phantom{CD} \\
    5  & \prot\prot\pip\pim\piz\piz & \leavevmode\phantom{0}300     & AB\phantom{CD}           \\

    6  & \prot\Lzero\Kp             & \leavevmode\phantom{00}54.5   & A\phantom{BCD}           \\
    7  & \prot\Lzero\Kp\pip\pim     & \leavevmode\phantom{00}20     & A\phantom{BCD}           \\
    8  & \prot\Szero\Kp             & \leavevmode\phantom{00}23.5   & A\phantom{BCD}           \\
    9  & \prot\Szero\Kp\piz         & \leavevmode\phantom{00}20     & A\phantom{BCD}           \\
    10 & \prot\Szero\Kp\pip\pim     & \leavevmode\phantom{000}2     & A\phantom{BCD}           \\

    11 & \prot\Lzero\Kp\piz         & \leavevmode\phantom{00}43     & AB\phantom{CD}           \\
    12 & \Dp\Kp\Lzero               & \leavevmode\phantom{00}64     & \leavevmode\phantom{A}B\phantom{CD} \\
    13 & \neut\Lzero\Kp\pip\piz     & \leavevmode\phantom{00}20     & \leavevmode\phantom{A}B\phantom{CD} \\
    14 & \Dz\pip\Kp\Lzero           & \leavevmode\phantom{00}30     & \leavevmode\phantom{A}B\phantom{CD} \\
    15 & \prot\Lzero\Kp\piz\piz     & \leavevmode\phantom{00}10     & \leavevmode\phantom{A}B\phantom{CD} \\
    16 & \prot\Lzero\Kp\piz\piz\piz & \leavevmode\phantom{000}7     & \leavevmode\phantom{A}B\phantom{CD} \\
    17 & \prot\Szero\Kzs\pip        & \leavevmode\phantom{000}9     & \leavevmode\phantom{A}BC\phantom{D} \\

    18 & \prot\prot\pip\pip\pim\pim & \leavevmode\phantom{0}227     & \leavevmode\phantom{AB}CD           \\
    19 & \prot\Lzero\Kzs\pip        & \leavevmode\phantom{00}30     & \leavevmode\phantom{AB}CD           \\
    20 & \prot\Lzero\Kp\pip\pim     & \leavevmode\phantom{00}21     & \leavevmode\phantom{AB}CD           \\
    21 & \neut\Lzero\Kzs\pip\pip    & \leavevmode\phantom{00}10     & \leavevmode\phantom{AB}C\phantom{D} \\
    22 & \prot\prot\Kzs\Kzs         & \leavevmode\phantom{000}1.6   & \leavevmode\phantom{AB}CD           \\

    \bottomrule
  \end{tabularx}
\end{table}

Background contributions to real photon hyperon decays (denoted by A in \cref{tab:bkg_cs}) were selected to contain \prot\Kp\Lzero in the final state (channels 6-11). Additional multipion channels (1 and 5) account for \pip misidentification as \Kp.

The main background contributions to hyperon Dalitz-decays (denoted by B in \cref{tab:bkg_cs}) originates from channels containing \Lzero and a dilepton source (mainly \piz Dalitz-decays, channels 9 and 11).
An important source of background in the higher dilepton invariant mass region (\textit{i.e.} $M_{\text{\dalitzp}}>M_{\text{\piz}}$) is related to \textDelta{} Dalitz-decay, associated with \Lzero production (channels 7 and 9).
The background originating from Dalitz-decays of non-strange resonances without a \Lzero in the final state but emitted together with \prot\pim pair (demanded in the \Lzero reconstruction) was estimated by contributions from \textDelta$^{+,0}$ because of their relatively large branching ratio to dileptons.
\Css for these channels were approximated by the corresponding data for multi-pion production, assuming that pions originate only from the \textDelta{} decays (channels 2 and 4).
This approach is taken to be an upper limit for this background contribution.

Combinatorial background in the dielectron measurements is determined from Dalitz-decays and/or external conversion of photons from neutral pions. Hence, production channels with one or multiple \piz were also included in the simulations.

\begin{figure*}[h]
  \centering
  \includegraphics[width=1.0\linewidth]{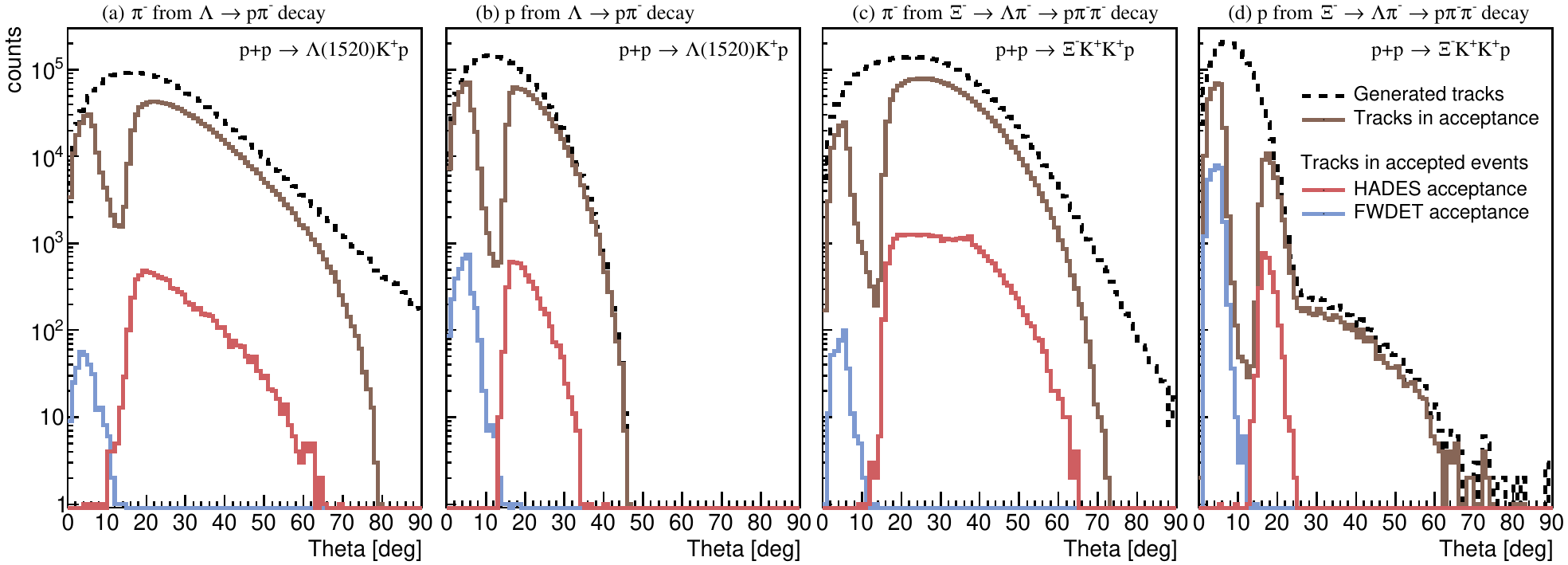}
  \caption{Polar angle distribution of the (a) pions and (b) protons emitted from \Lstard decay following the \pp\decays\prot\Kp\Lstard(\Lzero\vg) reaction, and (c) pions and (d) protons emitted from \Xim decay following the \pp\decays\prot\Kp\Kp\Xim reaction. The black dashed curve shows the distribution for all corresponding particles created in the collisions. The brown line shows the inclusive distribution for those emitted within either the \hades or the \fwdet acceptance. The red and the blue lines correspond to particles in the \hades and \fwdet acceptance, respectively, which are registered within fully reconstructed events \textit{i.e.} all required tracks of the full event are within the acceptance of \hades or the \fwdet.}
  \label{fig:angles}
\end{figure*}

The most abundant background channels for the \Xim (denoted by C in \cref{tab:bkg_cs}) are related to multi-pion production (channels 17 to 22), which have much larger production \css than the corresponding signal channels.
Fortunately, they can be effectively suppressed by applying a condition on the displaced \prot\pim decay vertex of the \Lzero.
Therefore, the important background channels contain a \Lzero hyperon and at least one \pim meson.
A special case for this class of reactions contains pions produced at displaced vertices w.r.t. the primary interaction vertex.
This situation can occur when a \Lzero is produced associated with a \Kzs meson (channels 17, 19, 21 and 22). The D group related to the \dilambda production contains the same background channels as the remaining \prot and \pim in the final state can contribute to additional \Lzero.

Existing \cs data for the background channels without strangeness were taken wherever possible from the compilation in Ref. \cite{L-B}.
If no data on a given channel are available at our beam energy, the results from higher energies were used, therefore the assumed \css provide an upper estimate for the expected background.
\Css for most of the strangeness-containing background channels were based on measurements of different channels at various beam energies collected from \cite{Flaminio:1984gr} and fit with a phase space distribution according to:
\begin{equation*}
  \sigma(E) = a \cdot \left(1 - \frac{E_{0} +2m_{p}}{E+2m_{p}}\right)^{b} \cdot \left(1 - \frac{E_{0}}{E+2m_{p}}\right)^{c}\text{,}
\end{equation*}
where $E$ is the kinetic energy of the beam particle, $E_{0}$ is the threshold beam energy for the production of the channel and $m_{p}$ is the proton mass.

\section{Simulation and analysis techniques}\label{sec:sim}

The selected benchmark and background channels have been simulated using the Pluto Monte-Carlo event generator \cite{Frohlich:2007bi} and processed with GEANT3 to model the detector acceptance. The simulations were performed for a \prot beam with \SI{4.5}{\gev} incident on a \SI{4.7}{\cm} long \ce{LH_2} target.

Simulated events were reconstructed with the same methods as for experimental data using the hydra software framework \cite{HADES}. The effect of the trigger applied during data taking was studied using a dedicated emulator, as described below.

\subsection{Acceptance for hyperon decays}
\label{sec:acc_hyp}

The daughter baryon from hyperon decay will be strongly forward peaked in the laboratory reference frame as a consequence of three main effects:
\begin{enumerate*}[label=\emph{(\alph*})]
  \item the produced hyperons have an anisotropic angular distribution in the center of momentum reference frame (c.m.)
  \item the decay kinematics of the hyperons and
  \item the boost of the final state particles into the laboratory frame resulting from the fixed target kinematics.
\end{enumerate*}
The kinematic boost is the most important of these three effects for the given set of reaction channels and beam energy.

The angular distribution in the c.m. reference frame of \Lzero produced at a beam kinetic energy of $E_\text{kin} = \SI{3.5}{\gev}$ shows about a factor two enhancement for forward emitted hyperons relative to $\theta_{c.m.}=\ang{90}$ \cite{PhysRevC.95.015207}.
This asymmetry is however expected to be significantly smaller for the production of the heavier  \Shype{}$^*$/\Lzero{}$^*$ states, \Xim and \dilambda in this study, which are produced at smaller excess energies  $\epsilon\sim$ \SIrange{0.2}{0.5}{\gev}, and thus is neglected  below.

The daughters from the hyperon decays relevant for these studies include a proton and either one pion for \Lzero\decays\prot\pim or two pions for the cascade, which proceeds through two consecutive weak decays: \Xim\decays\Lzero\pim and \Lzero\decays\prot\pim.
The large mass ratio between the proton and pion, together with the relatively small decay phase space for the hyperons result in the proton being emitted in a direction very close to the hyperon direction.
In contrast, the pion(s) will be emitted over a much wider angular range relative to the direction of the hyperon.
This effect is visible in \cref{fig:angles} (a) and (b) for the pion and proton daughters of \Lstard decay, respectively.
Similarly, the pion and proton daughters from \Xim decay are shown in frames (c) and (d), respectively.
The dashed black lines show the distributions of the particles emitted in the full solid angle, and the solid (brown) lines show the corresponding distribution for particles measured within the acceptance of either \hades or the new \fwdet.
Although most of the protons are emitted forward, a tail to larger angles is visible in frame (d), that results from decays after rescattering of the \Xim on the target or the detector materials.
The characteristic dip around \ang{10} is due to the lack of acceptance in the region between the new \fwdet and the existing \hades due to the magnet support ring.
The blue and red solid curves in \cref{fig:angles} show the corresponding distributions for accepted particles in the \fwdet and \hades, respectively, for those events in which all final state particles are registered in the detector acceptance.
The distributions clearly demonstrate that most pions are emitted into the \hades acceptance, whereas the protons are emitted preferentially into the forward direction with a large fraction detected in the \fwdet.

The quantitative importance of the new \fwdet for hyperon reconstruction close to the production threshold can be seen from the following results: \SI{88}{\percent} of events with all daughters of the \Xim inside the acceptance of \hades or the \fwdet have both pion tracks in the \hades acceptance and the proton track in the \fwdet.
Only \SI{11}{\percent} of \dilambda events are completely inside the \hades acceptance and the remaining \SI{89}{\percent} include at least one particle in the \fwdet.
The \fwdet detects \SI{41}{\percent} of all accepted protons tracks for the \Lstard decays.
The smaller acceptance gain for the \Lstard by adding the \fwdet compared to the \Xim is a consequence of the higher excess energy for this reaction ($\epsilon=\SI{0.5}{\gev}$).
The increased acceptance by a factor of two due to the \fwdet is sizeable and important for reconstruction of rare Dalitz-decays.
These simulations can be taken as a lower estimate of the detector acceptance, because any additional forward peaking of the production differential \cs in the c.m. frame will lead to a further increase of the \fwdet acceptance.

\subsection{Trigger filtering}
\label{sec_trigger}

An efficient and selective trigger condition is a key factor for the planned experiments. A fast trigger applied in \hades uses multiplicity information based on the number of hits in both of the time of flight META (Multiplicity Electron Trigger Array) detectors: TOF and RPC. For this work the same trigger conditions are applied that were used during the \prot+\prot experiment at \SI{3.5}{\gev} investigated by \hades in 2007. The multiplicity trigger used in that experiment was set to at least three hits in META (so-called "M3" trigger). The M3 trigger seems to be an optimal trigger for the planned measurements of hyperons because it strongly suppresses background from two-prong events and at the same time does not significantly reduce the number of reconstructed hyperon yield in the \hades acceptance.

Dalitz-decays of hyperon resonances \hyped \decays \Lzero{}(\prot\pim)\dalitzp produce four particles that need to be reconstructed.
Simulation studies of \Lstard decays show that the M3 trigger condition does not significantly reduce the number of reconstructed events inside the \hades acceptance. \cref{fig:angles} shows that either all four tracks are observed in \hades (\SI{59}{\percent}) or three in \hades and one (proton) in the \fwdet (\SI{41}{\percent}).

Reconstruction of \Xim decay requires the proton from the secondary \Lzero decay to be detected in the \fwdet and both pions to be detected in \hades. This situation occurs with \SI{88}{\percent} probability, thus, one of the associated kaons must also be detected in \hades in order to fulfill the M3 condition. Note that most spectator protons are detected in the \fwdet, similar to protons from \Lzero decay. The simulations show that the M3 trigger condition reduces the number of reconstructed \Xim events by only \SI{10}{\percent}.

For the \dilambda reconstruction both pions from \Lzero decays are emitted into \hades and the two protons either both enter the \fwdet or one of them enters the \fwdet and the second goes through \hades. In this case the M3 trigger reduces the number of reconstructed \dilambda events by only \SI{3}{\percent}, as most of the time at least three particles hit the HADES.

\subsection{Track reconstruction and particle identification}

\begin{figure}[!t]
  \centering
  \includegraphics[width=1.0\linewidth]{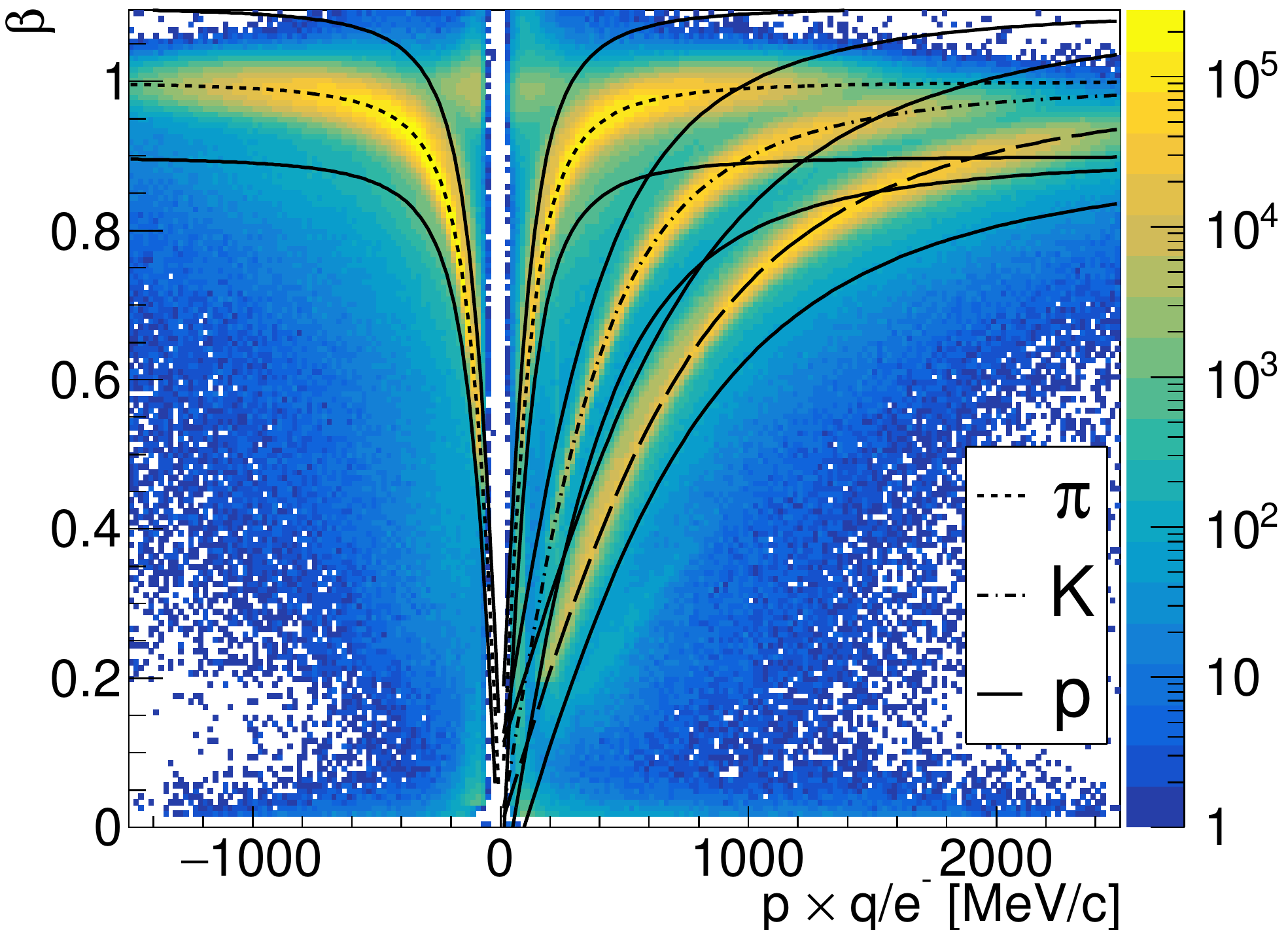}
  \caption{Particle identification based on the $\beta$ vs momentum correlation measured in \hades. The dashed, dot-dashed and dotted lines designate the $\beta_c$ reference for protons, kaons and pions, respectively. The solid lines denote the $\beta_\text{c}\pm 0.075$ selection range for each particle species.}
  \label{fig:pid_cuts}
\end{figure}

The particle tracking in \hades is based on the four planes of the multi-wire drift chambers (MDC) with each plane consisting of 6 wire layers. Two planes are located in front of the toroidal magnetic field and the other two behind the field, allowing the track deflection to be measured (see \cref{fig:fd_integration}). Each pair of planes is positioned in a magnetic field-free region, where  straight track segments are reconstructed.
The track segments before and after the magnetic field volume are connected by a curved track section. The momentum of the particle is calculated from the curvature of the track in the magnetic field (see \cite{Agakishiev:2009am} for details).
Next, the reconstructed track is matched with an associated hit in the TOF or RPC detectors.
The reconstructed path length and the ToF information are combined to calculate the particle velocity ($\beta c$) and conditions on the correlation between the momentum and the velocity can be used to differentiate between particle species of different mass.
Depending on the simulated channel, the particle identification (PID) methods applied slightly differ.
For the \Xim and \dilambda analyses, the PID selection is based on selection bands calculated by varying the expected velocity (expressed as $\beta_\text{c}$) for a given mass hypothesis and momentum, as shown by the solid lines in \cref{fig:pid_cuts}.
For the hyperon Dalitz-decays, the PID is determined by calculating the mass squared from the measured $\beta$ and momentum.
Particles with squared masses in the range \SIrange{40}{240}{\mevsc} and \SIrange{650}{1127}{\mevsc} are used to identify \textpi{} and \prot, respectively.
Leptons are identified by matching the reconstructed track with a spatially correlated ring measured in the RICH (for details see \cite{Agakishiev:2009am}). For the reconstruction of real photon decay of hyperons, the PID was based on a neural network, as described in \cref{sec:hypreal}.

\begin{figure}[!b]
  \centering
  \includegraphics[width=1.0\linewidth]{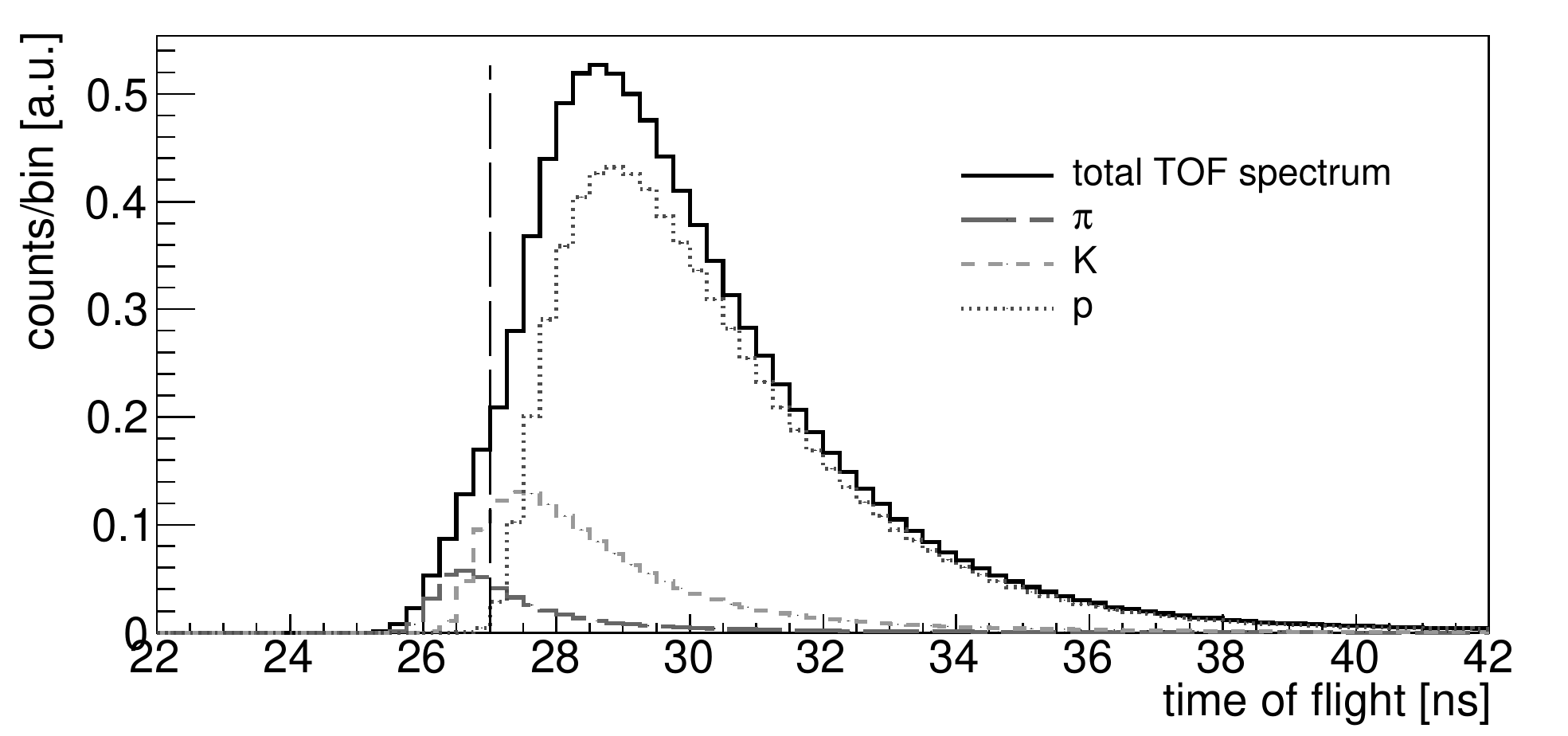}
  \caption{ToF spectra for different particle species measured in the FRPC, produced in the reaction \pp\decays\prot\Kp\Kp\Xim.}
  \label{fig:xim_rpc_tof}
\end{figure}

Unlike the \hades detector, the \fwdet is positioned entirely in a region with no magnetic field, thus no direct momentum measurement is possible. Instead, for a given mass hypothesis the momentum of a track can be deduced by combining the flight path information of the \fwdet with the ToF measurement in the FRPC.
Although the measured ToF is not sufficient to completely separate protons from pions and kaons  by itself (see \cref{fig:xim_rpc_tof}),  it is  nevertheless sufficient to provide an important constraint for various event hypotheses, since the path lengths have very little dispersion.
For example, by demanding that ToF $>$ \SI{27}{\nano\second}, as indicated by the dashed line in \cref{fig:xim_rpc_tof},  the pion contribution is suppressed by \SI{45}{\percent} compared to only \SI{0.7}{\percent} for the proton yield.

These studies show that protons from hyperon decays are emitted preferentially in the forward direction and the \fwdet plays an important role in their detection.
Hence, for the \Xim and \dilambda reconstruction the event hypothesis requires that all particles in the \fwdet are protons, and that both pions are in the \hades acceptance and identified as pions based on the PID methods explained above.
For the hyperon Dalitz-decays, a pair of leptons and a pion must be found in \hades, and the remaining daughter proton is either identified in \hades or reconstructed in the \fwdet.

\subsection{Inclusive \Lzero reconstruction} \label{sec:la_reco}

Reconstruction of the \Lzero is common to all analyses described here. As a neutral particle with the dominant \prot\pim weak decay it can be identified by its displaced pion-proton vertex ($c\tau=\SI{7.89}{\cm}$). As discussed in the previous section, the event hypothesis applied in the analysis assumes that the particle detected in the \fwdet is a proton and the \pim track candidate is selected based on the PID in \hades with a high (\SI{>90}{\percent}) purity. For \SIrange{20}{40}{\percent} of events (depending upon which reaction is studied) both daughter particles are emitted into the \hades acceptance and can be easily identified.

\begin{figure}[!b]
  \centering
  \includegraphics[scale=0.65]{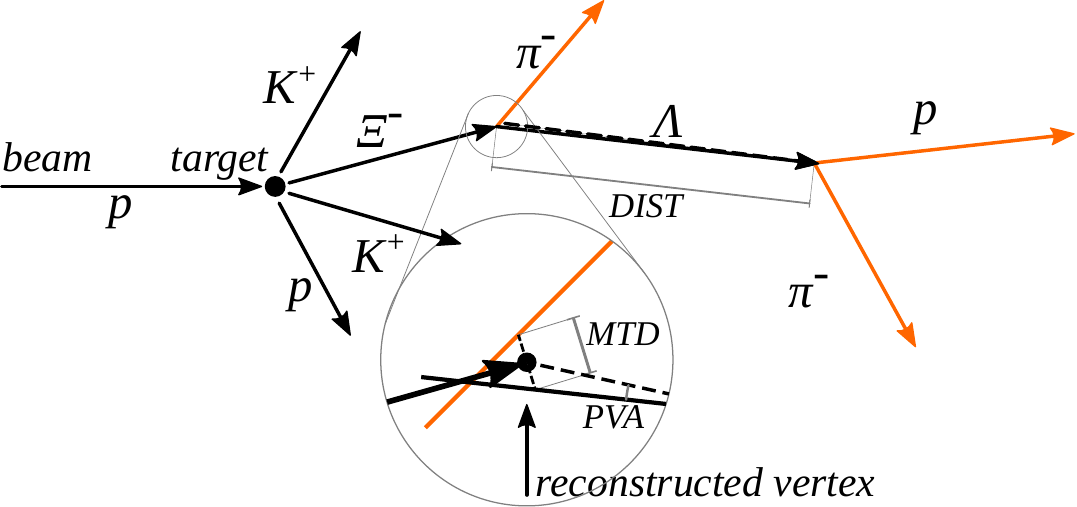}
  \caption{\Xim decay topology and the derived observables. The dots indicate the reconstructed primary vertex and the \Xim decay vertex.
    The gray lines represent the distance (DIST) between the decay vertices and the minimal track distance (MTD) between the daughter tracks of a given vertex. The Pointing Vector Angle (PVA) is the angle between the reconstructed \Lzero momentum vector and the vector connecting the reconstructed \Xim and \Lzero decay vertices.
    The orange tracks show particles registered in \hades or the \fwdet and used in the signal reconstruction.}
  \label{fig:xim_topo}
\end{figure}

\begin{figure*}[!t]
  \centering
  \includegraphics[width=1.0\linewidth]{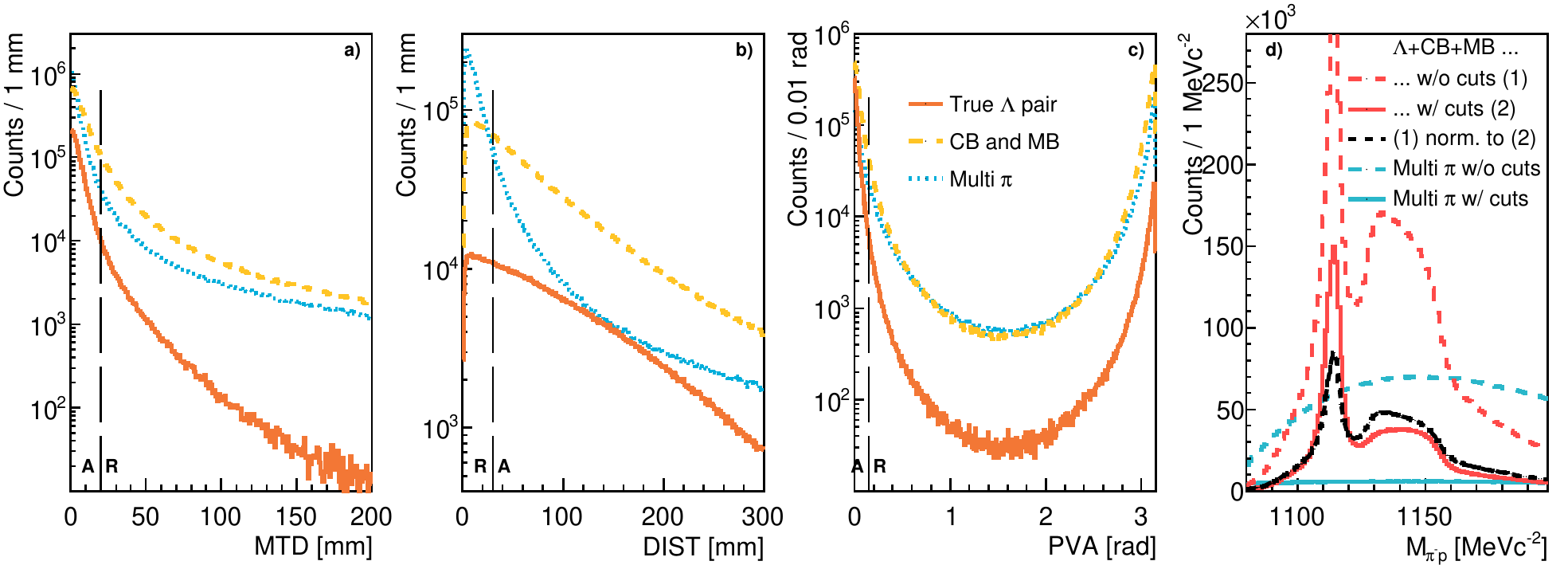}
  \caption{Examples of the MTD (a), DIST (b) and PVA (c) selections (vertical dashed lines) for the \pp\decays\Xim\Kp\Kp\prot reaction. The effect of these selections on the \Lzero invariant mass distribution is shown in (d). The solid orange line is for the true \Lzero, the dashed yellow is CB and MB from the same reaction, and the dotted blue is the distribution for the background stemming from multi-pion production \pp\decays2\prot2\pip2\pim. The vertical line denotes the selections applied in the analysis: the letters A and R denote the accepted and rejected regions, respectively. In (d), the red line is the reconstructed signal from the \Xim production channel, and the blue line is the distribution for multi-pion production. The dashed line is with no selections applied and the full line is after applying all three selections (a--c). The black dashed line is the \Xim channel without selections, normalized to the same integral as after the selections. The enhanced \sbs ratio is clearly visible.}
  \label{fig:lambda_xim_cuts}
\end{figure*}

However, a small fraction of events with \Lzero decays features a pion registered in the \fwdet and a proton in \hades or even both particles detected in the \fwdet. These cases give rise to a small misidentification background (MB), due to the incorrect particle hypothesis in the \fwdet.
There is also a combinatorial background (CB) that arises from combining true protons and pions that do not originate from the decay of a single \Lzero.
Furthermore, there is physics background (PB) originating from other channels that contain a \Lzero in the final state, but have a much larger production \cs than the signal channel (see \cref{tab:bkg_cs}).
Finally, there is also a background originating from combinations of true protons and pions stemming from multi-pion production channels that do not include a \Lzero.
The rejection strategies adopted to suppress the contributions of these background sources to the final spectra are presented below.

The characteristic delayed decay of the hyperon is used to discriminate the signal from many of these background sources.
The typical topology for \Lzero decay following \Xim decay is presented in \cref{fig:xim_topo}. This is a more general production scheme than the \Lzero produced in the primary vertex, and illustrates well the observables used for the CB and MB suppression.

In the first step a pion and proton are identified and their four-vectors are combined to obtain the four-vector of their mother particle (\Lzero candidate). The Minimal Track Distance (MTD) between the proton and pion is used to identify a true track pair from the \Lzero decay, for which MTD should ideally be equal to zero. However, the limited track resolution results in a distribution of this observable for track pairs originating from the same vertex. A selection on this variable alone does not adequately discriminate between emission from the primary and the displayed vertices. Nevertheless, it allows to significantly reduce CB arising from tracks originating from different vertices.
Based on the simulations a maximum value for MTD was found to be in the range \SIrange{15}{25}{\milli\metre} and is further adjusted for the specific reaction channel to optimize the signal significance \signs, where $S$ is the total yield of signal events and $B$ is the integrated background yield within $\pm 3\sigma$ of the signal peak. \Cref{fig:lambda_xim_cuts} (a) shows the distribution of MTD for true \Lzero daughter pairs in the \Xim production channel (solid orange line), CB and MB backgrounds from the same final state (dashed yellow line), and random pairs from multi-pion background channels (dotted blue line).

\begin{figure}[!b]
  \centering
  \includegraphics[scale=0.65]{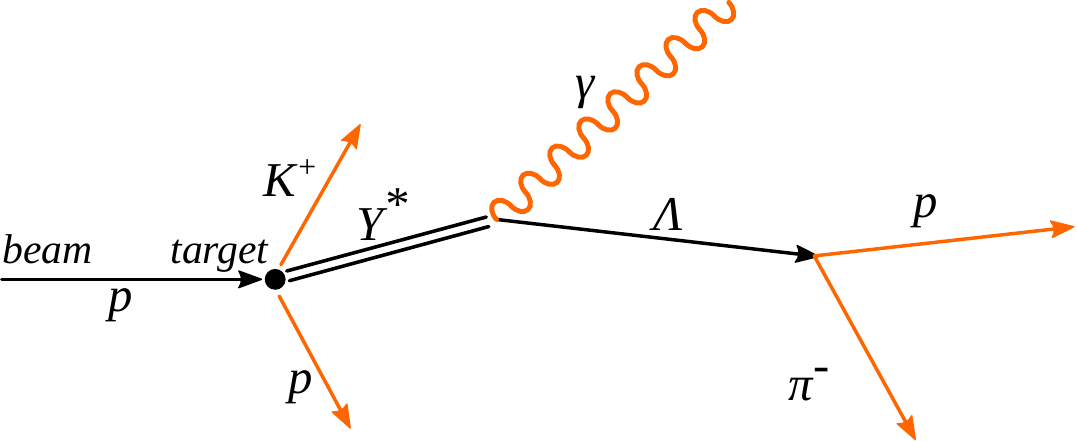}
  \caption{Topology of hyperon radiative decay. The orange tracks indicate the particles registered in \hades or the \fwdet and are used in the event reconstruction.}
  \label{fig:ygamma_topo}
\end{figure}

The \Lzero vertex distance (DIST) measures the distance between the \Lzero production vertex and the displaced decay vertex. The primary vertex is located in the \ce{LH_2} target, which extends from \SIrange{-55}{-8}{\mm} along the beam axis.
Requiring a minimum value of DIST reduces the contamination of track pairs resulting from the primary vertex as well as contributions from background channels with \Kzs in the final states, which decays into \pip\pim ($c\tau=\SI{2.68}{\cm}$), where the \pip can be misidentified as a proton.
Although this selection discards a fraction of the \Lzero decays close to the primary vertex, the improved \sbs ratio is more important, leading to a higher significance for the channel when using this selection. The simulations indicate that the optimal minimum value for DIST should be in the range of \SIrange{25}{35}{\milli\metre} to select \Lzero events.
An example of this distribution for the \Xim reaction channel is shown in \cref{fig:lambda_xim_cuts} (b).

The third discriminating observable used in these simulations is the Pointing Vector Angle (PVA), defined as the angle between the reconstructed 3-momentum vector of the \Lzero candidate and the line segment joining the reconstructed production and decay vertices. The optimal selection on this distribution is in the range \SIrange{0}{1}{\radian}. It depends on the accuracy of the \Lzero vector reconstruction and the vertex determination, and is correlated with the DIST observable. An example of this selection for the \Xim channel is shown in \cref{fig:lambda_xim_cuts} (c). The second peak near PVA = \textpi{} appears for cases where DIST is small and due to the finite resolution the secondary vertex is reconstructed before the primary vertex ($z_S < z_P$) even though the \Lzero is emitted in the forward direction ($p_z > \SI{0}{\mevsc}$).

\Cref{fig:lambda_xim_cuts} (d) shows the composite effect of these selections (solid lines) on the reconstructed \prot\pim invariant mass distribution for the \Xim signal (red) and the multi-pion background channel (blue). The mass distribution for the \Xim reaction without these selections (dashed red line) is also scaled to the integral of the distribution after the selections (black dashed). The comparison of the red to the black dashed distributions allow to appreciate the improved \sbs ratio.

\section{Results}\label{sec:results}

All simulations used a full cocktail of signal and background channels that are listed in \cref{tab:sig_cs,tab:bkg_cs}. An arbitrary number of events were simulated, and then weighted according to the corresponding \cs.

\subsection{Exclusive real photon decay of hyperons}\label{sec:hypreal}

\begin{figure}[!t]
  \includegraphics[width=0.99\linewidth]{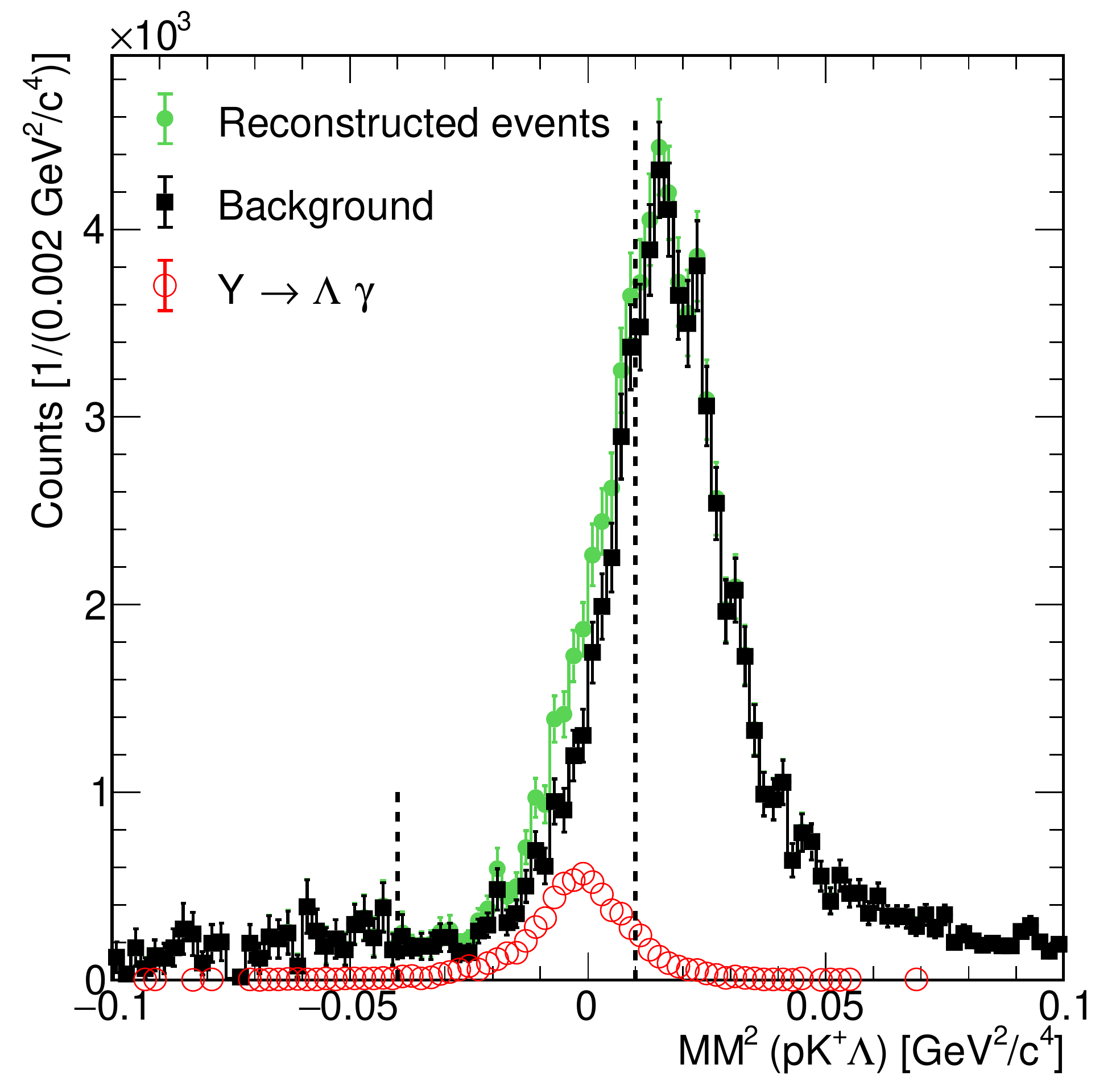}
  \caption{Distribution of the \prot \Kp \Lzero squared missing mass for reconstructed events in the analysis of the Lambda radiative decay. The red open circles and black squares show the projected signal and background contributions, respectively. The green circles show the sum. The vertical dashed lines mark the signal region.}
  \label{fig:MM2pkL}
\end{figure}

The topology for \hyped\decays\Lzero\photon is shown in \cref{fig:ygamma_topo}. In contrast to the other analyses presented in this paper, events with real photon decay of hyperons must be reconstructed exclusively in order to suppress the large background of \hyped\decays\Lzero\piz, where one of the \piz decay photons is not detected.

The first step in the reconstruction of this channel is the selection of events with the correct final state charged particle candidates, \textit{i.e.} \prot, \prot, \Kp and \pim.
In order to obtain very pure lists of particle candidates, deep-learning algorithms with the capability to model complex and non-linear data dependencies are employed here for PID.
An Artificial Neural Network (ANN), implemented within the PyTorch framework \cite{paszke2017automatic}, is used.
The network is used as a multi-class classifier to distinguish among five particle species \prot, \pip, \pim, \Km and \Kp.
The input layer consists of seven neurons corresponding to the number of features, which are the three momentum components, the energy loss in the MDC and TOF, as well as the charge and velocity of the particles.
The output layer consists of six neurons, five correspond to the particle species and the additional neuron allows the network to classify other particle types as ''other particle''. The network is trained on the following final states: \prot\prot\pip\pim\piz, \prot\prot\Kp\Km and \prot\Kp\Lzero\piz.
The best validation accuracy was obtained by sequentially combining three fully connected layers.
In addition, a \SI{50}{\percent} dropout to each layer was applied to prevent the model from over-fitting. The network converged quickly and has a classification accuracy of \SI{99}{\percent}, \SI{97}{\percent}, \SI{94}{\percent}, \SI{97}{\percent} and \SI{93}{\percent} for \prot, \Kp, \pip, \pim and \Km, respectively.

For each event, the point of closest approach of the primary (direct) proton \pprot and the \Kp candidate was defined as the primary vertex. Since there is more than one proton per event, the MTD between each proton and the \Kp tracks is calculated. The proton and corresponding primary vertex with the smallest MTD is selected. Furthermore the primary vertex is required to be located within the target volume ($\SI{-60}{\mm} < z < \SI{-5}{\mm}$, and $r < \SI{6}{\mm}$).

\begin{figure}[!t]
  \includegraphics[width=0.99\linewidth]{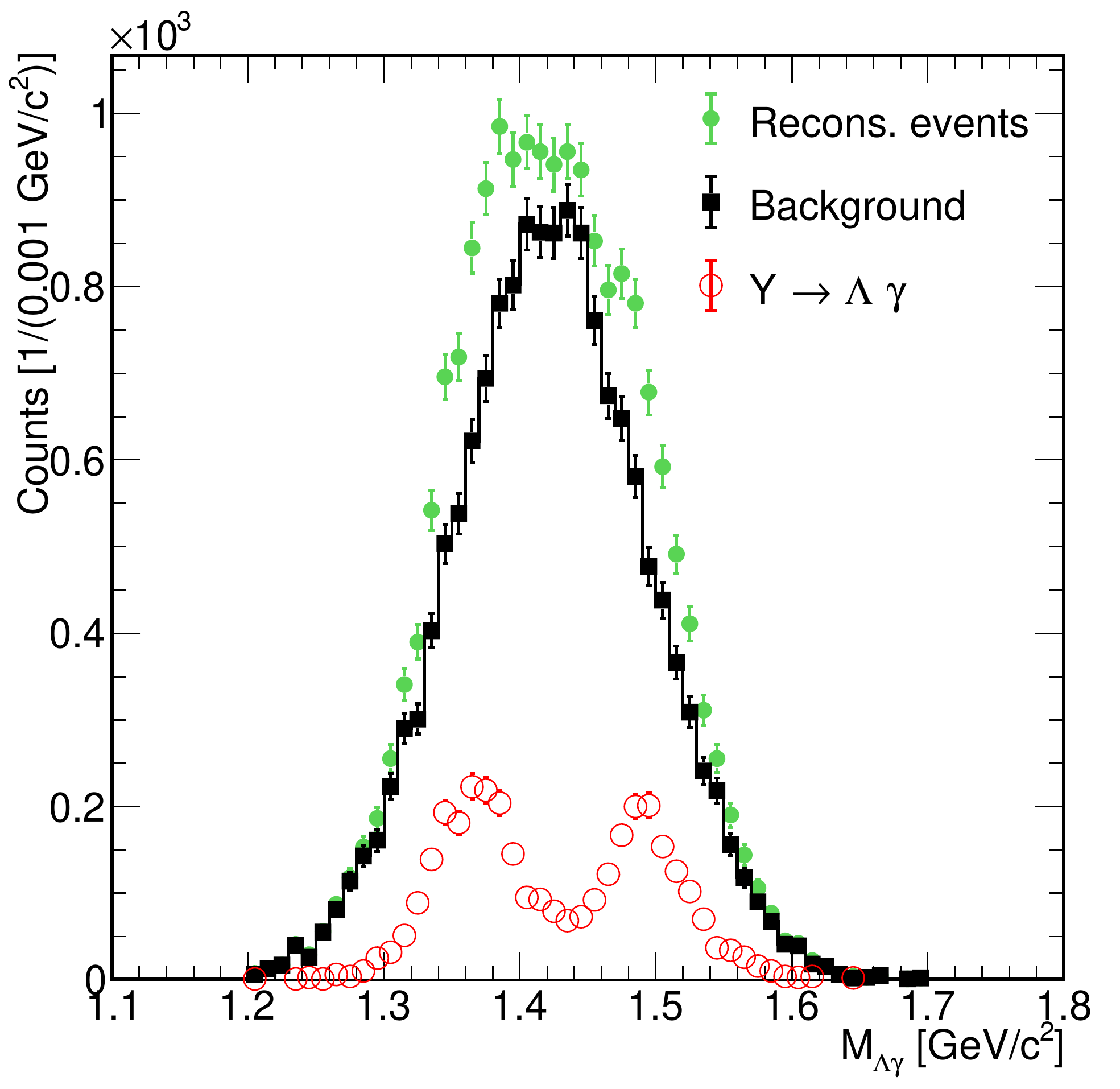}
  \caption{Projected \Lzero \photon invariant mass distribution of the reconstructed events in the signal region (the green circles). The black squares are background channels and the red open circles is the Y \decays \Lzero \photon signal after background subtraction. The contributions are scaled according to the expected integrated luminosity corresponding to 28 days of measurement with the \ce{LH_2} target.}
  \label{fig:HyperonMass}
\end{figure}

\begin{figure}[!b]
  \centering
  \includegraphics[scale=0.65]{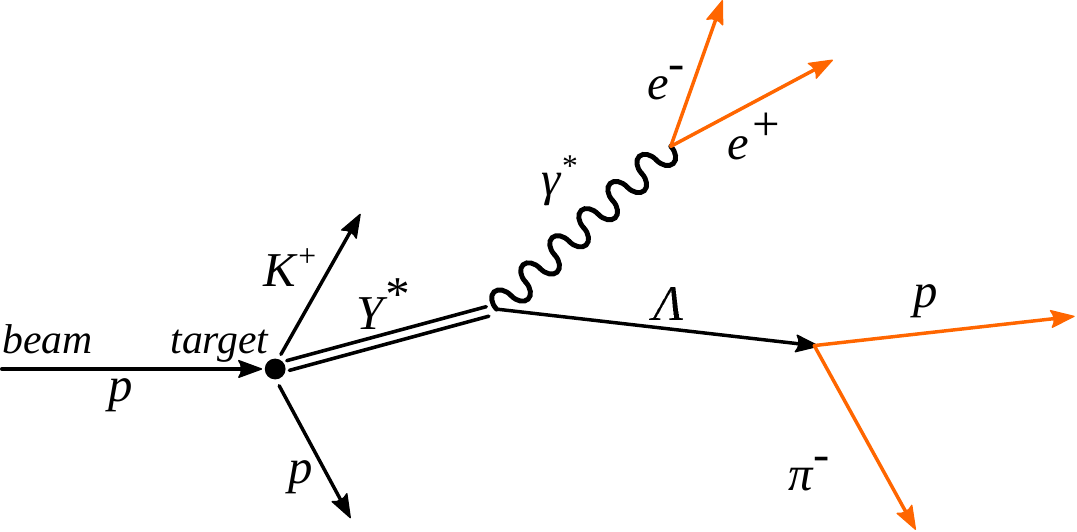}
  \caption{Topology of hyperon Dalitz-decay. The orange tracks indicate the particles registered in \hades and used in the reconstruction.}
  \label{fig:yepem_topo}
\end{figure}

\begin{figure*}[!t]
  \centering
  \begin{minipage}[b]{0.49\textwidth}
    \includegraphics[width=1.0\linewidth]{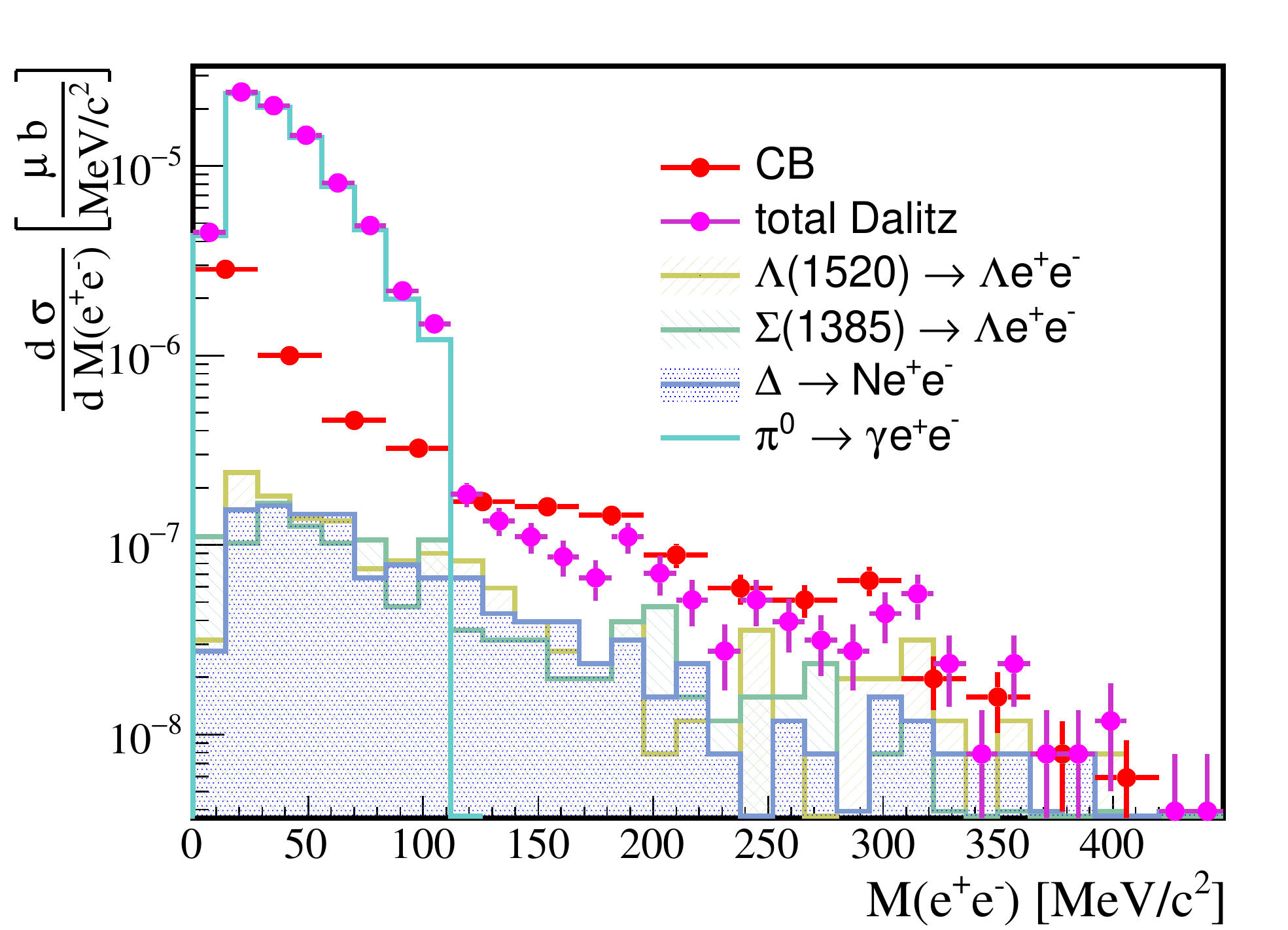}
    \centering (a)
  \end{minipage}
  \hfill
  \begin{minipage}[b]{0.49\textwidth}
    \includegraphics[width=1.0\linewidth]{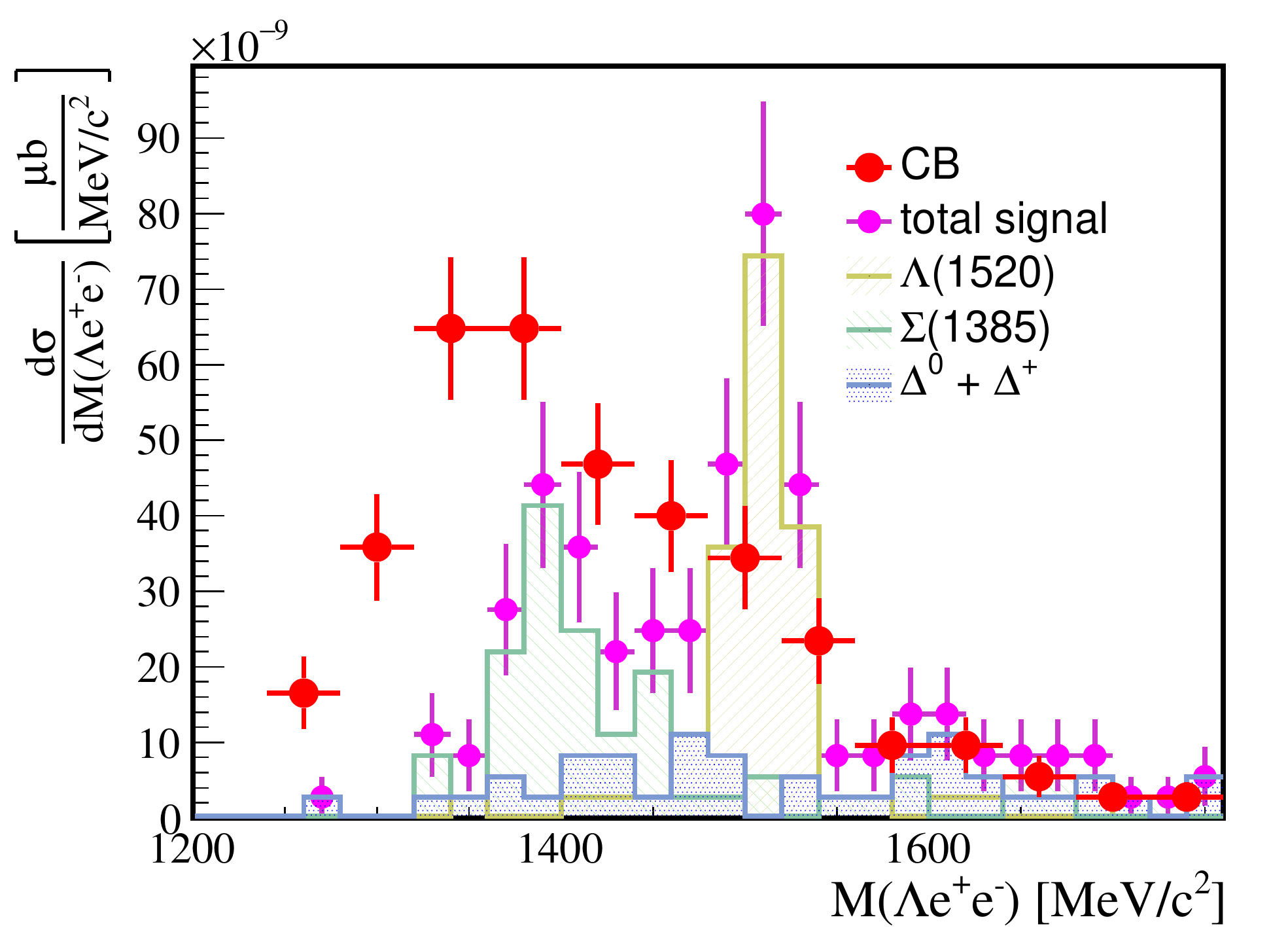}
    \centering (b)
  \end{minipage}
  \caption{Dilepton invariant mass spectrum from signal \vg and background \piz sources (a) and reconstructed \Lstard, \Sstar and \Lstar peaks (b), see text for details. The statistical uncertainty corresponds to a four week measurement at a luminosity of \SI{1.5e31}{\per \centi\meter\squared \per \second}.}
  \label{fig:lstard_reco}
\end{figure*}

The intermediate \Lzero has been reconstructed by demanding that its decay products (\textit{i.e.} \prot and \pim) originate from a delayed vertex. All combinations of \prot and \pim candidates are used to form \Lzero candidates. For each \Lzero candidate the decay vertex is defined as the Point of Closest Approach between the two tracks.
If both tracks are in \hades, then the MTD between the \prot and \pim tracks is required to be smaller than \SI{15}{\mm}. If a track is measured in the \fwdet it is assumed to be a proton and the MTD is required to be less than \SI{20}{\mm}.
The boost of the \Lzero into the laboratory reference frame ensures that the \Lzero and its decay proton are almost collinear, while the \pim should have a larger deviation from the direction of the \Lzero momentum vector.
Thus, the Distance of Closest Approach between the \prot track and the primary vertex is required to be smaller than for the \pim track and the primary vertex.
Furthermore, the PVA is required to be smaller than \SI{0.5}{\radian}.

Photon candidates are identified as an energy cluster in the ECAL above a certain threshold.
The contribution from charged particles is suppressed by requiring the cluster to be spatially uncorrelated with the RPC hits.
The 4-vector of the candidate photon is calculated from the cluster energy and the cluster position \cite{czyzycki2011electromagnetic}.
Photon candidates are required to have $0.96<\beta<1.04$, and an energy deposition above \SI{0.35}{\gev}.

Since several background channels mimic the signal, two kinematic variables have been used for further background suppression:
The \pprot\Kp squared missing mass is required to be in the range $1.6 < \mathit{MM}^2(\text{\pprot\Kp}) < \SI{2.6}{\sgevsc}$.
The squared \pprot\Kp\Lzero missing mass ($\mathit{MM}^{2}(\text{\pprot\Kp\Lzero}))$ should peak at zero for signal events, since only one photon is missing. Thus it is required to be in the range $-0.04 < \mathit{MM}^{2}(\text{\pprot\Kp\Lzero}) < \SI{0.01}{\sgevsc}$.
The signal range is asymmetric about zero in order to reduce contamination from the \prot\Kp\Lzero\piz background, which is peaked at the mass squared of the \piz.
The results are shown in \cref{fig:MM2pkL}, where the red open circles indicate the signal, the black squares the background, and the sum of both is shown as the green circles.
The vertical dashed lines indicate the region of the squared missing mass selected for the signal.

Finally, the excited hyperon is reconstructed in the \Lzero \photon invariant mass distribution and presented in \cref{fig:HyperonMass}. The background is mostly from the \prot\Kp\Lzero\piz channel  (reaction 11 in \cref{tab:bkg_cs}).
The slight shift in the peak position of the \Sstarz and \Lstard results from the response function of the ECAL.
The estimated overall acceptance times reconstruction efficiency for \Sstarz, \Lstar and \Lstard is \SI{0.030}{\percent}, \SI{0.030}{\percent} and \SI{0.026}{\percent}, respectively.
The corresponding significance to reconstruct each of these states is of 16, 0.3 and 15.

\subsection{Dalitz-decay of excited hyperons}

The Dalitz-decay of hyperons (\cref{fig:yepem_topo}) was reconstructed in the reactions where the primary hyperon resonance decays into a \Lzero and a virtual photon \vg, which then decays into an \dalitzp dilepton pair. The inclusive reconstruction of \Lzero candidates proceeded similar to the \Xim channel. The main difference is that the decays of the hyperon resonances occurs at the primary vertex. The subsequent \Lzero decay appears at a displaced vertex. Therefore, the \dalitzp pair originates at the primary vertex located in the target. The z-coordinate of the secondary vertex is required to be \SI{>0}{\mm}, for the given target position extending from \SIrange{-55}{-8}{\mm}. Furthermore, a MTD \SI{<20}{mm} was demanded for \Lzero candidates to reduce the background from uncorrelated pion-proton pairs. The minimal opening angle for the dilepton pair is \ang{4} to reduce conversion background, which is mostly emitted at lower opening angles.

\Cref{fig:lstard_reco} (a) shows the reconstructed \dalitzp invariant mass spectrum. The combinatorial background (CB) originating from uncorrelated \dalitzp pairs is shown by the red dots. The magenta dots represent the sum of all reconstructed Dalitz \dalitzp signal. The blue-green histogram represent \dalitzp pairs that originate from \piz decays.
Dalitz-decays of the \textDelta{} are denoted by the blue histogram. The yellow and green histograms show the spectra originating from hyperon (\Lstard and \Sstar, respectively) Dalitz-decays.
The figure clearly shows that the region of invariant mass below \SI{140}{\mevsc} is dominated by the \piz Dalitz decay, which can not be fully suppressed by the conditions on the \Lzero reconstruction.
Above the \piz range, the main background comes from the \textDelta{} Dalitz decay.

\Cref{fig:lstard_reco} (b) shows the \Lzero\dalitzp invariant mass distribution, where the dilepton mass is required to be above the \piz mass (\textit{i.e.} $M_\text\dalitzp > \SI{140}{\mevsc}$). Clear peaks from the \Lstard and \Sstarz are visible above a broad background from the \textDelta. The low branching ratio for the \Lstar results in too little yield to be measured here.
The product of the acceptance times reconstruction efficiency is estimated to be about \SI{0.48}{\percent} and \SI{0.58}{\percent} for the \Sstarz and \Lstard, respectively.

These simulations were performed under the assumption that the decaying particles are point-like.
As discussed in the introduction, it is expected that the form factors will enhance the decay rate in the high mass region and consequently increase the count rates with respect to these simulations.
The expected count rates are summarized in \cref{tab:expected_cr}.

\subsection{Inclusive \Xim production}

\begin{figure}[!t]
  \centering
  \includegraphics[width=1.0\linewidth]{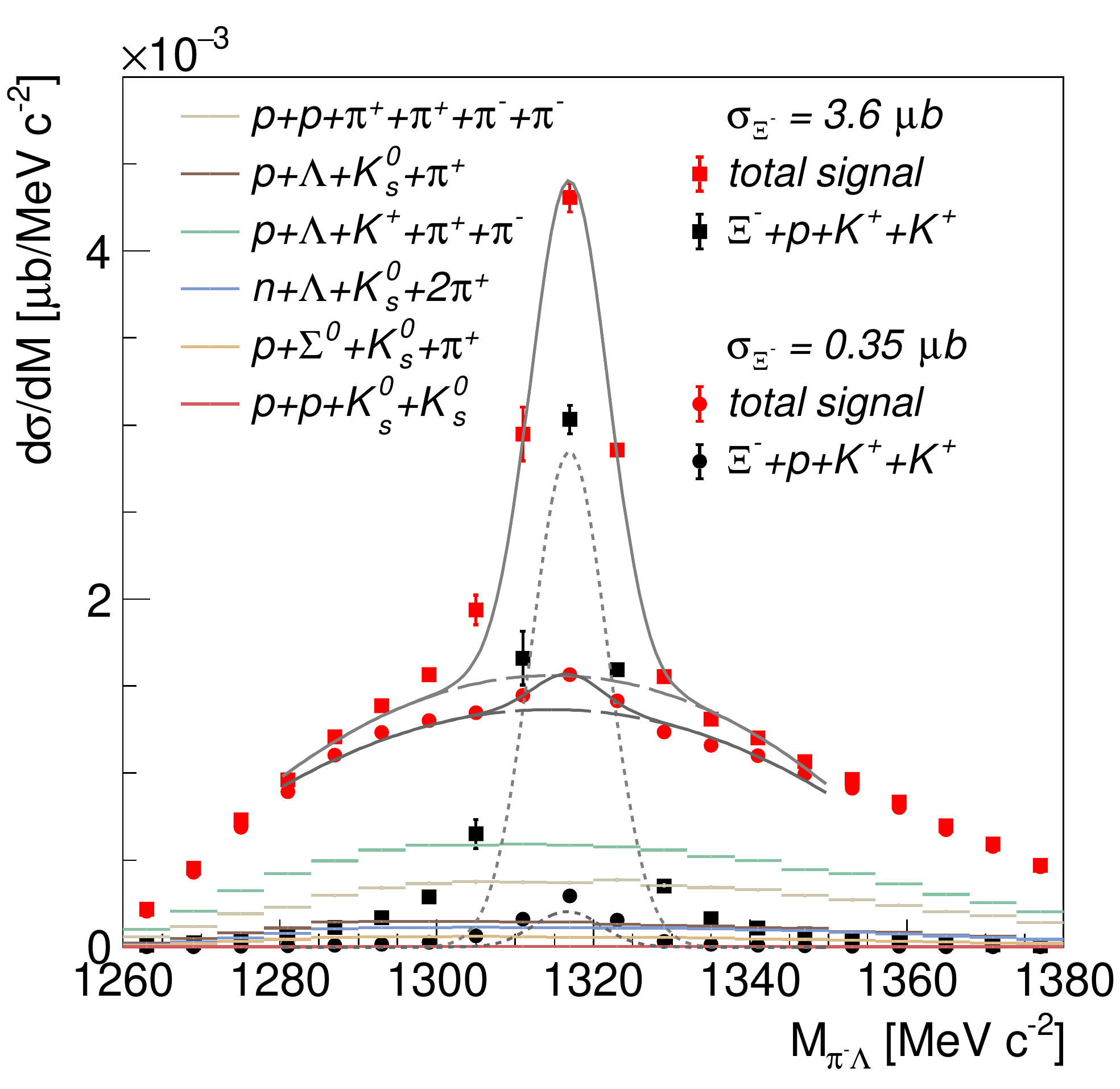}
  \caption{Reconstructed \Lzero\pim invariant mass spectrum. The red points indicate the distributions for a \Xim signal \cs of \SIlist{3.6;0.35}{\mub}. The black points indicate the corresponding distributions after the combinatorial background (CB) has been subtracted.}
  \label{fig:xim_reco_xim}
\end{figure}

\begin{figure}[!t]
  \centering
  \includegraphics[width=1.0\linewidth]{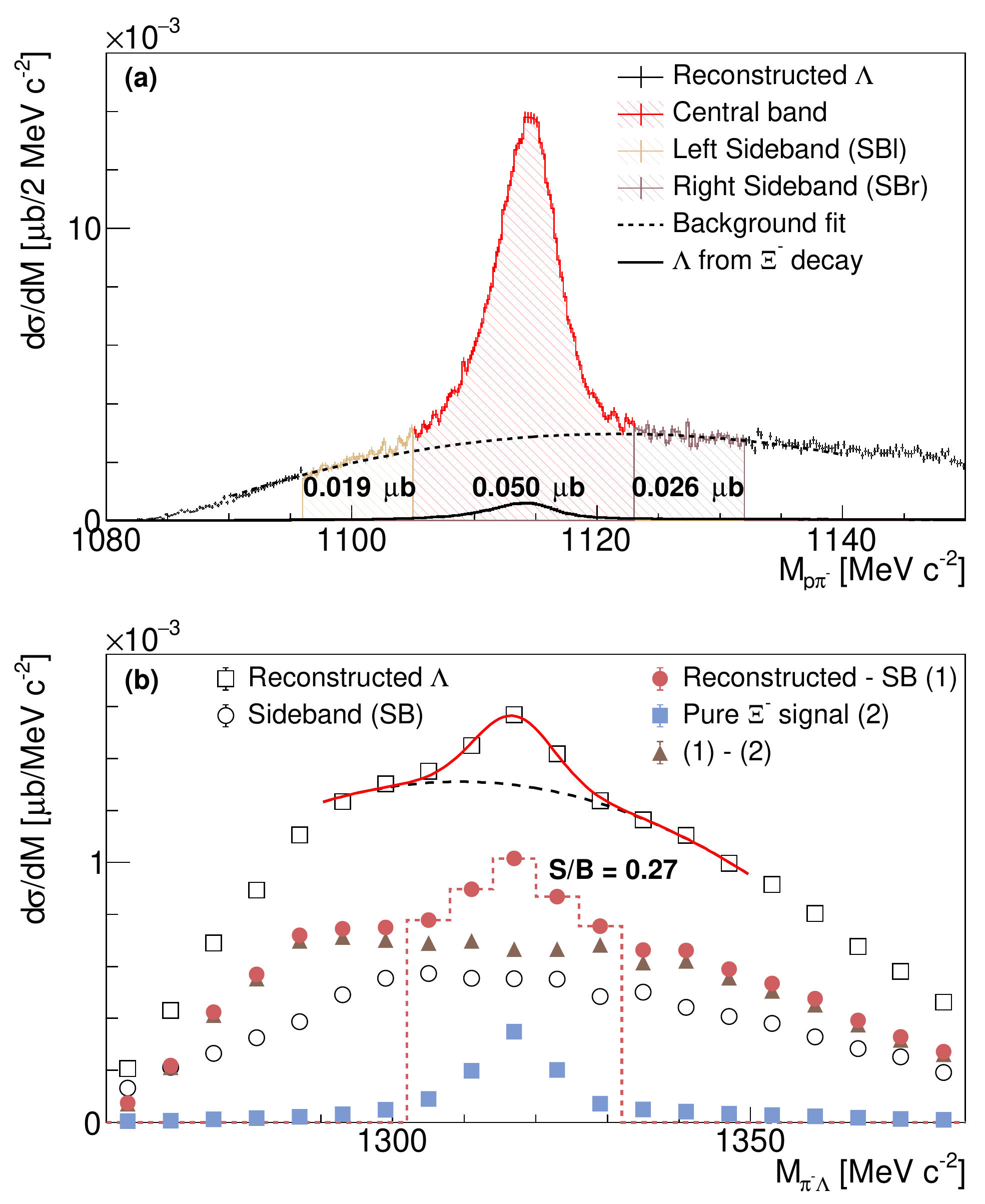}
  \caption{Sideband analysis for the \Xim signal. (a) Sideband regions for the daughter \Lzero invariant mass spectrum. The dashed line shows a fit to estimate the background under the signal peak. The numbers show the integrated \cs values of the background for each background region. (b) \Lzero\pim invariant mass spectra. The open square symbols show the distribution for all signal candidates. The open circles represent the spectrum for the \Lzero sideband region, and the solid circles are the full results after subtracting the sideband region. The solid square symbols show the \Xim signal for reference. The brown triangles mark the estimated background line shape.}
  \label{fig:xim_sb}
\end{figure}

The \Xim decay was reconstructed by the following chain of weak decays: in the first step the \Xim decays into a \Lzero\pim pair and in the second step the \Lzero decays into a \prot\pim pair. The \Lzero was reconstructed as described above, considering pion tracks from \hades and proton tracks identified in either \hades or the \fwdet. The \Xim was reconstructed combining \Lzero candidates and pions from \hades. In order to effectively reduce the combinatorial background from the signal and the misidentification background, the following set of topological selections has been applied (\cref{fig:xim_topo}):
\begin{enumerate*}[label=\emph{\alph*})]
  \item MTD$_\text\Lzero$ for the proton and pion track candidates from \Lzero decays is required to be $<\SI{25}{mm}$,
  \item the z-coordinate for the \Lzero decay vertex is required to be in the range $\SI{-20}{mm} < z_\text\Lzero < \SI{300}{mm}$,
  \item MTD$_\text\Xim$ between the \Lzero and the pion track candidates from \Xim decay is required to be $<\SI{20}{mm}$,
  \item the PVA of \Lzero track to the line between \Xim and \Lzero decay vertices is lower than \SI{0.15}{\radian}.
\end{enumerate*}
An additional selection on the squared missing mass of the \Xim candidate relative to the beam+target system $MM^2 > \SI[parse-numbers=false]{2050^2}{\smevsc}$ has been applied to suppress background channels.
The specific selection values have been chosen to optimize the significance.
To reduce misidentifying pions and kaons as protons in the \fwdet, only those tracks with $t_\mathrm{tof} > \SI{27}{\nano\second}$ (\cref{fig:xim_rpc_tof}) are accepted.
The overall acceptance times reconstruction efficiency for the \Xim is determined to be about \SI{1.68}{\percent}.

\begin{figure}[!b]
  \centering
  \includegraphics[scale=0.65]{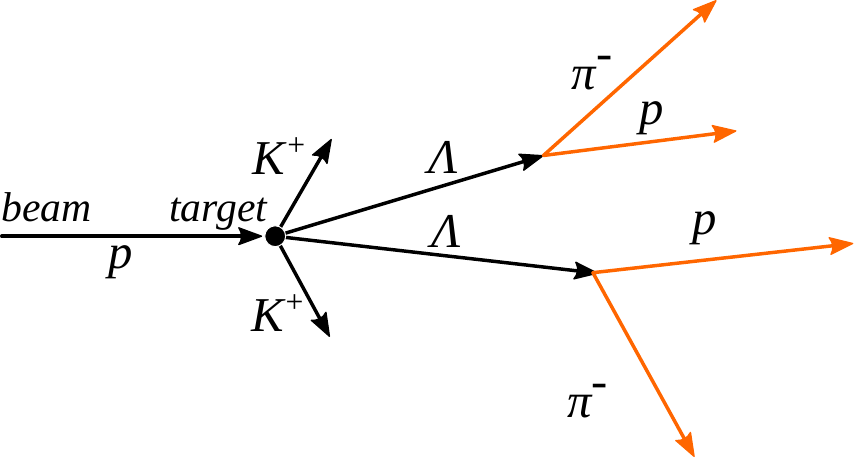}
  \caption{\dilambda decay topology. The orange tracks indicate the particles registered in \hades and used in the reconstruction.}
  \label{fig:lala_decay}
\end{figure}

\begin{figure*}[!t]
  \centering
  \includegraphics[width=1.0\linewidth]{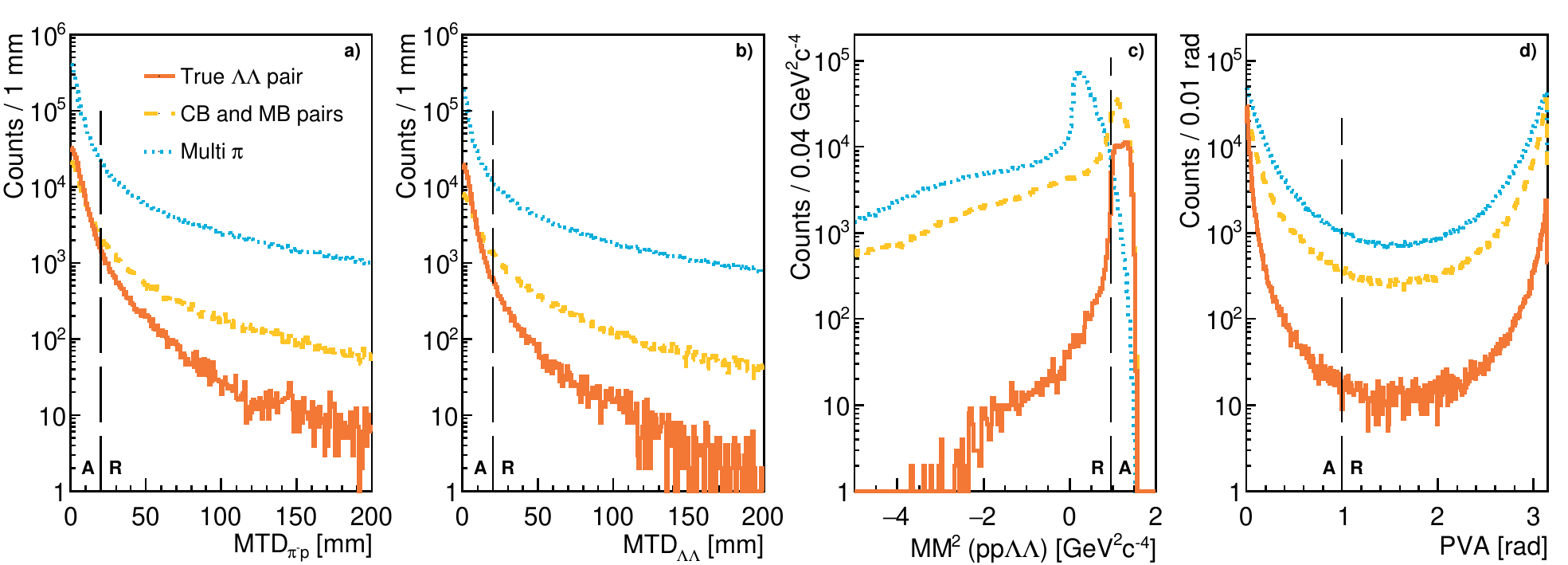}
  \caption{The MTD$_\text{\pim\prot}$
    distribution for \Lzero daughters in \dilambda events (a), the MTD$_\text{\dilambda}$
    distribution for the two \Lzero candidates (b), the squared missing mass MM$^2$ value (c) and the PVA values for both \Lzero candidates(d). The vertical lines mark the accepted and rejected regions, denoted by the letters A and R, respectively.}
  \label{fig:ll_cuts}
\end{figure*}

\begin{figure}[!b]
  \centering
  \includegraphics[width=1.0\linewidth]{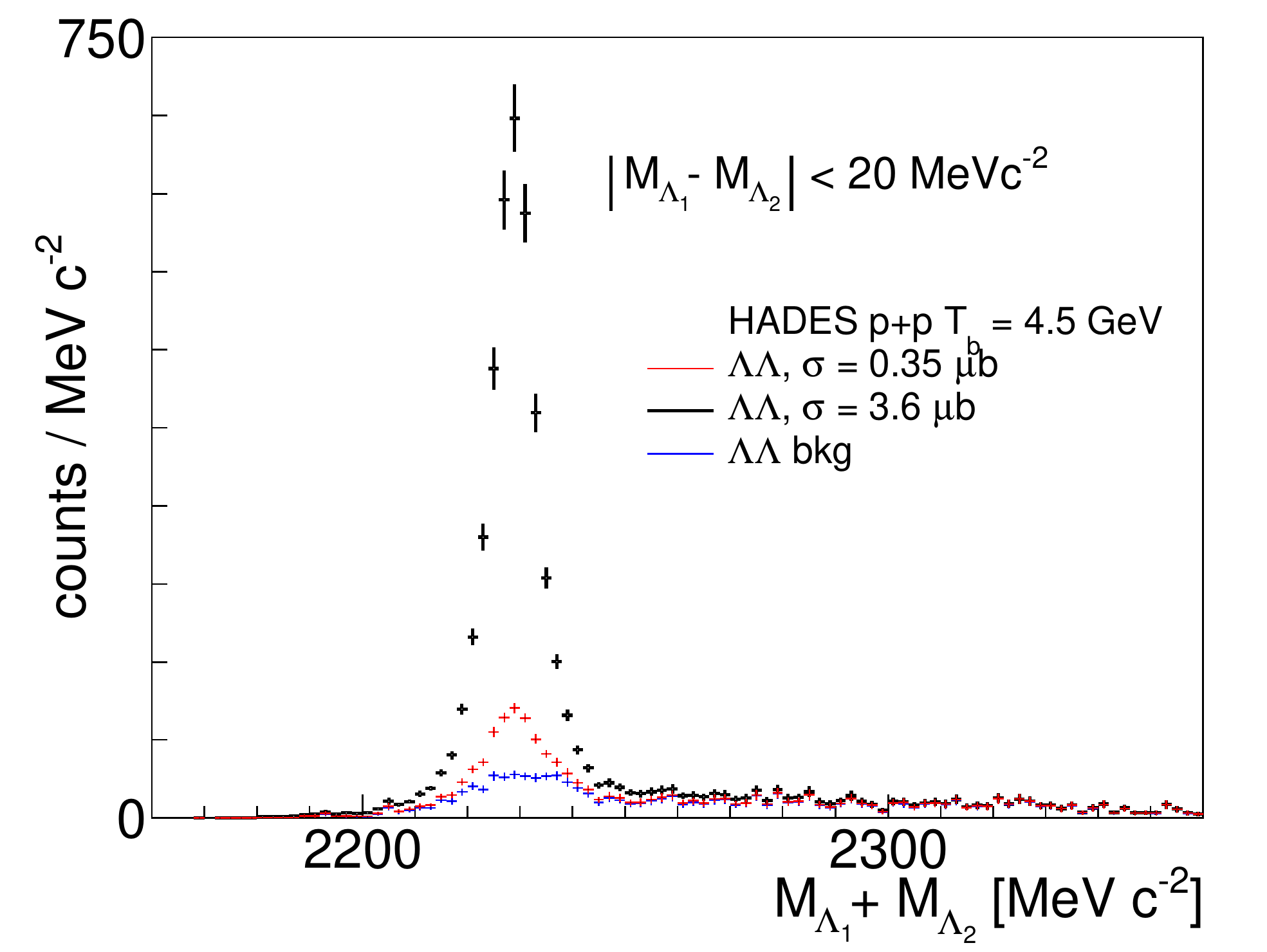}
  \caption{Sum of the invariant masses for both \Lzero candidates in \dilambda decay for the two \Lzero{s} mass difference less than \SI{20}{\mevsc}.}
  \label{fig:lala_mass_sum}
\end{figure}

Two values of the \Xim production \css (\SI{3.6}{\mub} and \SI{0.35}{\mub}) have been used for these simulations.
The resulting \Lzero\pim invariant mass distributions are shown in \cref{fig:xim_reco_xim}, together with various background channels (denoted by different colored lines).
For the case with the higher \cs (red squares) the \Xim peak is clearly visible on the top of the background. For the other case (red circles) the \Xim peak is less pronounced.
For the reference, the true \Xim signal is shown for both cases by the black symbols.
The full spectrum for each case was fit with the sum of a fifth order polynomial for the background and a Gaussian for the signal.
These fits are drawn as solid gray lines and the background contribution is drawn as long-dashed lines. The mass of the \Xim peak is found to be \SI{1317.77 +- 0.62}{\mevsc}.
It differs slightly from the PDG value of \SI{1321.71 +- 0.07}{\mevsc}. Also the peak of the pure \Xim distribution is asymmetric.
This asymmetry in the line shape and the shift of the reconstructed peak position result from \fwdet tracks which require further corrections, such as for energy loss in the detector material and the different velocities of the hyperon and its daughter proton. The statistics presented in this figure corresponds to \num{2.5} days of data taking with the \ce{LH_2} target.

\begin{figure}[!b]
  \includegraphics[width=1.0\linewidth]{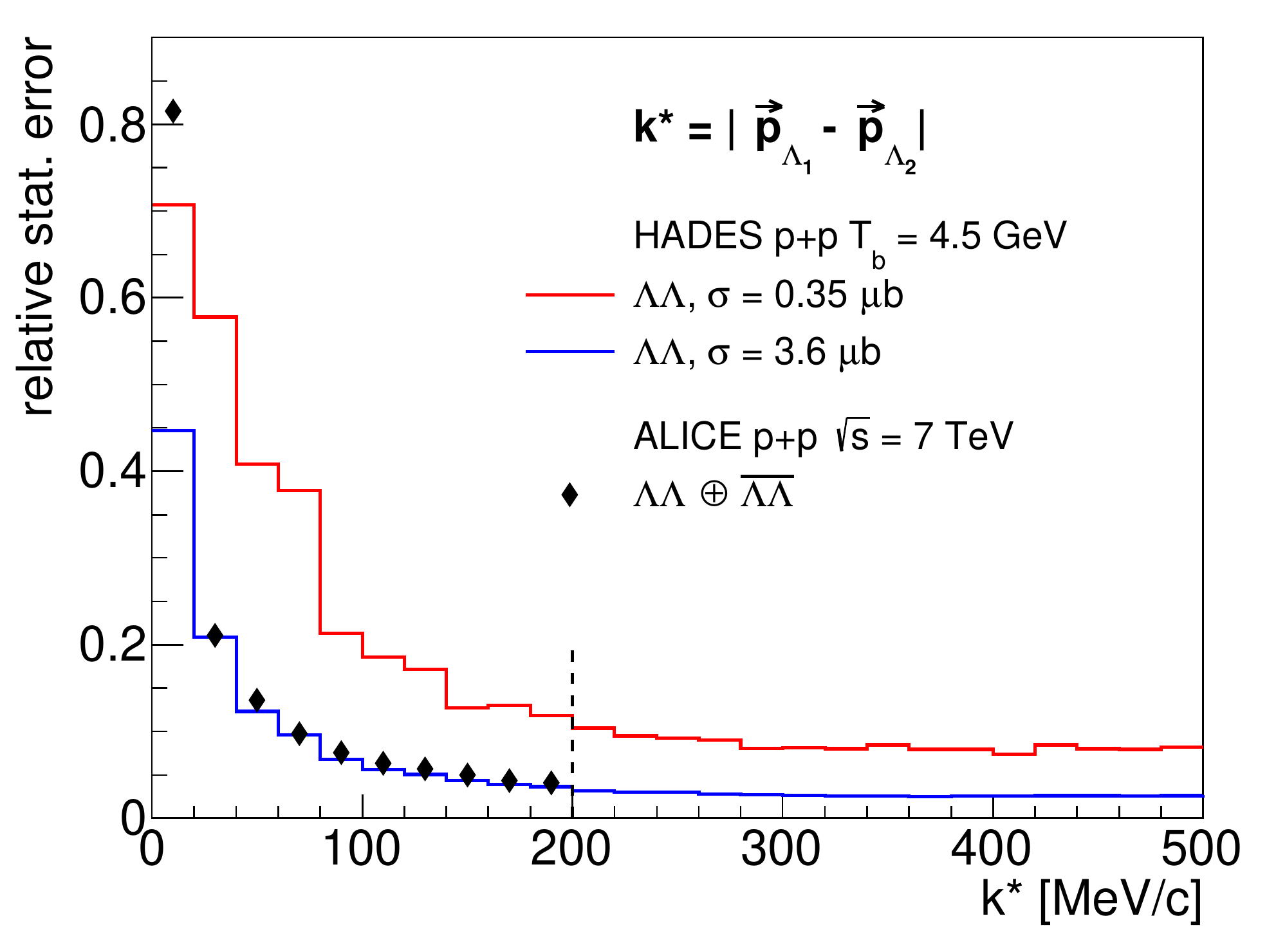}
  \caption{Relative error of the yield for \dilambda pairs as a function of the relative momentum.}
  \label{fig:LLcor}
\end{figure}

In order to reduce the background under the \Xim peak due to uncorrelated \pim\prot pairs, a sideband analysis of the \pim\prot invariant mass spectrum has been used (see \cref{fig:xim_sb} (a)).
This background is related to contributions from reactions without \Lzero production.
A region on the left \SIrange{1096}{1105}{\mevsc} and one on the right \SIrange{1123}{1132}{\mevsc} were used to estimate the background. The selected sidebands have the same width as the \Lzero peak region and their integrated yield is slightly smaller than the signal region due to the non-linear shape of the background .
The background yield under the peak was estimated from a fit, presented by the dashed black line.
Since only a small fraction of the \Lzero{s} in the peak originate from the \Xim signal (solid black line), only a fraction of the background under the \Xim signal can be removed by this sideband analysis.

\begin{table*}[!t]
  \caption{Expected count rates for luminosity $\mathcal{L} = \SI{1.5e31}{\per \centi\meter\squared \per\second}$ available using a liquid hydrogen (\ce{LH_2}) target and a polyethylene (\ce{CH_4}) target of the same dimensions. The branching ratios include a factor of 0.64 for each \Lzero\decays\prot\pim decay included.}
  \label{tab:expected_cr}
  \begin{tabularx}{1.0\textwidth}{L{1.0}L{0.6}L{0.6}L{0.6}L{0.6}L{0.6}}
    \toprule
    decay process                       & $\sigma [\si{\micro\barn}]$ & BR            & $\epsilon \cdot \mathrm{acc}$ [\si{\percent}] & $\sfrac{\text{counts}}{\text{day}}$ (\ce{LH_2}) & $\sfrac{\text{counts}}{\text{day}}$ (\ce{CH_4}) \\
    \midrule
    \Sstarz\decays\Lzero\dalitzp        & 56                          & \num{8.94e-5} & 0.48                                          & 15                                              & 105                                             \\
    \Lstard\decays\Lzero\dalitzp        & 69                          & \num{6.93e-5} & 0.58                                          & 18                                              & 126                                             \\
    \Lstard\decays\Lzero\pim\pip        & 69                          & \num{4.22e-2} & 1.4                                           & \leavevmode\phantom{0}\num{2.64e4}                         & \leavevmode\phantom{00}\num{1.85e5}                        \\
    \Sstarz\decays\Lzero\photon         & 56                          & \num{9.07e-3} & 0.030                                         & 99                                              & 692                                             \\
    \Lstard\decays\Lzero\photon         & 69                          & \num{7.03e-3} & 0.026                                         & 82                                              & 574                                             \\
    \Xim\decays\prot\pim\pim            & $\phantom{0}3.6$            & 0.64          & 1.68                                          & \leavevmode\phantom{0}\num{2.43e4}                         & \leavevmode\phantom{00}\num{1.70e5}                        \\
    \Xim\decays\prot\pim\pim            & $\phantom{0}0.35$           & 0.64          & 1.68                                          & \leavevmode\phantom{0}\num{2.43e3}                         & \leavevmode\phantom{00}\num{1.70e4}                        \\
    \prot\prot\decays\Lzero\Lzero\Kp\Kp & $\phantom{0}3.6$            & $0.64^2$      & 0.34                                          & \leavevmode\phantom{0}\num{3.15e3}                         & \leavevmode\phantom{00}\num{2.20e4}                        \\
    \prot\prot\decays\Lzero\Lzero\Kp\Kp & $\phantom{0}0.35$           & $0.64^2$      & 0.34                                          & \leavevmode\phantom{0}\num{3.15e2}                         & \leavevmode\phantom{00}\num{2.20e3}                        \\
    \bottomrule
  \end{tabularx}
\end{table*}

For each sideband the \prot\pim\pim invariant mass distribution was calculated just as for the \Lzero peak region. The sum of the resulting distributions (open circles in \cref{fig:xim_sb} (b)) was scaled to the yield of the background under the \Lzero peak (region below the dashed line in the central region in \cref{fig:xim_sb} (a)) and subtracted from the \prot\pim\pim invariant mass distribution for the \Lzero peak region (open squares). The resulting \Lzero--\pim invariant mass distribution is presented by the solid red circles. For comparison the true \Xim signal marked by the blue squares is also drawn.

The region marked by the dashed red line was then used to extract the \sbs ratio and the significance. The total yield in this mass region minus the pure \Xim signal (\sigs) (blue squares) defines the background (\bkgs) (solid triangles). The \sbs ratio is \num{0.27} and the estimated significance in this invariant mass region for one day of data taking is estimated to be 28, allowing for the \Xim to be reconstructed and the production \cs to be determined.

\subsection{\dilambda production}

The topology of double \Lzero production is shown in \cref{fig:lala_decay}.
Both \Lzero{s} are emitted directly from the primary reaction vertex.
The reconstruction procedure does not distinguish the \Lzero particles, but double counting is avoided, as described below.
Reconstruction of the \Lzero decays follows the standard \Lzero procedure described in \cref{sec:la_reco} with MTD$_\text{\pim\prot}$ for both \Lzero{s} set to \SI{20}{mm}.
To reduce CB and MB, three additional selections on the two \Lzero systems were applied:
\begin{enumerate*}[label=\emph{\alph*})]
  \item missing mass of (\prot{$_\text{beam}$}, \prot{$_\text{target}$}, \Lzero, \Lzero) $> \SI{980}{\mevsc}$,
  \item minimal track distance between both \Lzero{s} MTD$_\text{\dilambda} < \SI{20}{\mm}$ and
  \item the PVA was required to be less than \SI{1.0}{\radian}.
\end{enumerate*}
The effect of these selections is shown in \cref{fig:ll_cuts} for the \pp \decays \dilambda\rest as well as the CB, MB and multipion backgrounds.
The estimated overall acceptance times reconstruction efficiency for this channel is \SI{0.34}{\percent}.

Events with \Lzero pairs are selected by requiring at least two distinct \Lzero candidates in the event, and that the difference between their invariant masses is smaller than \SI{20}{\mevsc}. The distribution of the sum of the candidates masses is shown in \cref{fig:lala_mass_sum}, where both \cs   estimates investigated have been considered (black and red points). The common background for both cases is shown in blue. This background originates from events of reactions 17--20, 22 (\cref{tab:bkg_cs}) with one \Lzero and the second candidate is contained in the mass window by chance.

Events containing two \Lzero{}s open the possibility to study \dilambda correlations.
Based on the overall acceptance times reconstruction efficiency determined above
(\SI{0.34}{\percent}), we expect an integrated yield of 220--2200 reconstructed events per day, depending on the assumed production \cs.

The corresponding statistical precision for the relative momentum spectrum ($k^*$) is shown in \cref{fig:LLcor} for both \cs estimates and is compared to the published values from ALICE.
Information on the \dilambda interaction can be primarily gained in the low $k^*$ region. We expect 312 and 3120 events with $k^* < \SI[per-mode=symbol]{0.2}{\gevc}$ in a 4 week experiment for the two scenarios. The yield and statistical precision are expected to be similar to the results from ALICE for the upper bound \cs. More importantly, even for the lower bound \cs, the expected precision is better in HADES in the region with $k^* < \SI[per-mode=symbol]{20}{\mevc}$, which is most sensitive to the interaction parameters.
Systematic uncertainty due to the large freeze-out volume and feed--down effects are important limitations for the existing results and those effects are expected to be significantly smaller for HADES with \pp \ at $E_\text{kin} = \SI{4.5}{\gev}$.

\subsection{Count rate estimates}

The count rates for these studies have been calculated based on the following parameters:
\begin{enumerate*}[label=\emph{\alph*})]
  \item the reconstruction efficiency times acceptance for the various channels, as obtained from the simulation studies described above,
  \item a beam duty cycle of \SI{50}{\percent}, and
  \item two options for the maximum luminosity as described in \cref{sec_det}
\end{enumerate*}

\Cref{tab:expected_cr} summarizes all the channels considered. The production \css for the proton-proton reactions and the branching ratios are given in columns 2 and 3. The product of the acceptance and reconstruction efficiencies is given in column 4. Columns 5 and 6 give the expected count rates for the \ce{LH_2} and the PE targets, respectively.

\section{Summary}\label{sec:summary}

Dalitz decays of hyperons is an experimentally unexplored process, which can be studied for the first time by using the excellent capabilities of HADES to measure rare dilepton decays.
Such measurements will provide complementary information on hyperon structure and the role of strange quarks in baryons. The latter will be investigated by comparing these results to those transitions recently measured for non-strange baryons of same spin and parity (i.e \Sstarz vs $\Delta(1232)$ and $\Lambda(1520)$ vs $N^*(1520)$).

Production cross sections of the higher mass hyperon resonances investigated here are unknown in the energy range of future FAIR experiments.
Hence, the proposed measurements will provide an essential input for the understanding of the production mechanism in proton-proton interaction in this energy regime and will also establish an important reference for \prot+A and A+A collisions.
In particular, the production of double strangeness offers a unique opportunity to study strangeness production close to the threshold, where only few production channels are expected to contribute.
In this regime, characterized by a small relative momentum between the produced hadrons, final state interactions are expected to play a significant role. Therefore, hyperon-hyperon and nucleon-hyperon correlation studies are very promising tools to measure their interaction potential.

The feasibility studies presented here show that all the benchmark channels with hyperon Dalitz-decays, except \Lstar, can be measured in \hades including the new \fwdet, which has been described in detail in this paper. The biggest uncertainty in these results comes from scarce \cs data for various signal and background channels. Thus, a systematic error of \SI{40}{\percent} is assigned to the count rates for most channels. However, the  \cs estimates for the double strange baryons \Xim and \dilambda are sensitive to unknown factors and have much larger uncertainty, up to an order of magnitude. The predicted count rates for the electromagnetic decays of \Sstarz and \Lstard favor the use of the polyethylene target combined with a high luminosity operating mode that will become possible with an upgraded \hades DAQ. For all benchmark channels, four weeks of measurement time is considered sufficient to measure the corresponding \css.

Currently, the \hades collaboration discusses a possible extension of the tracking region to angles \SIrange{7}{18}{\degree} allowing to cover blind spots of the spectrometer. Furthermore, a fast trigger hodoscope in front of the inner MDC is considered for a better on-line event selection. Such improvements will result in a further significant extension of the acceptance, as seen in \cref{fig:angles}, and better trigger possibilities.

\begin{acknowledgement}

  We gratefully acknowledge financial support by the Polish National Science Centre, grant 2016/23/P/ST2/04066 POLONEZ funded from the European Union's Horizon 2020 research and innovation program under the Marie Skłodowska-Curie grant agreement No. 665778.
  \begin{figure}[!ht]
    \vspace{-3ex}
    \centering
    \includegraphics[width=0.35\linewidth]{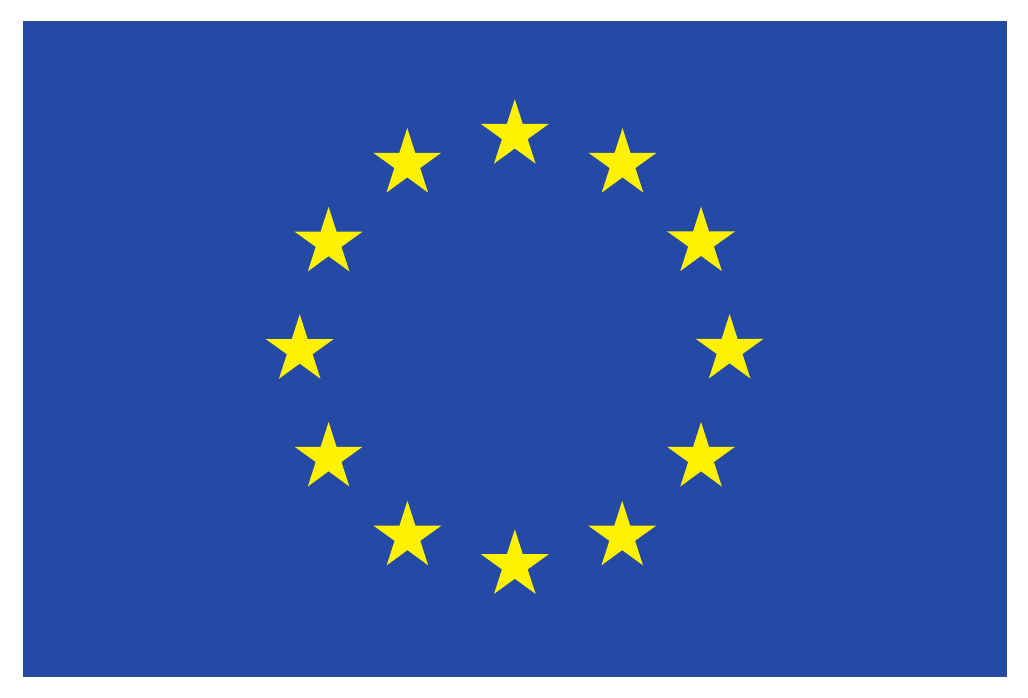}
  \end{figure}
  \vspace{-3ex}

  \noindent
  The collaboration gratefully acknowledges the support by SIP JUC Cracow, Cracow (Poland), National Science Center, 2017/25/N/ST2/ 00580, 2017/26/M/ST2/00600;
  Forschungszentrum J\"ulich, J\"ulich (Germany);
  TU Darmstadt, Darmstadt (Germany), VH-NG-823, BMBF: 05P18RDFC1, DFG GRK 2128, DFG CRC-TR 211;
  Goethe-Univers\-ity, Frankfurt (Germany) and TU Darmstadt, Darmstadt (Germany), ExtreMe Matter Institute EMMI at GSI Darmstadt;
  TU M\"{u}nchen, Garching (Germany), MLL M\"unchen, DFG EClust 153, GSI TMLRG\-1316F, BMBF 05P15WOFCA, SFB 1258, DFG FAB898/ 2-2;
  NRNU MEPhI in the framework of the Russian Academic Excellence Project (contract No. 02.a03.21.0005, 27.08.2013), RFBR funding within the research project no. 18-02-40086, Ministry of Science and Higher Education of the Russian Federation, Project "Fundamental properties of elementary particles and cosmology" No 0723-2020-0041,
  JLU Giessen, Giessen (Germany), BMBF:05P12RGGHM;
  IPN Orsay, Orsay Cedex (France), CNRS/IN2P3;
  NPI CAS, Rez, Rez (Czech Republic), MSMT LM2018112, OP VVV CZ.02.1.01/0.0/0.0/16013/0001677, LTT17003.
\end{acknowledgement}

\printbibliography

\end{document}